\documentclass[journal=nalefd,manuscript=letter]{achemso}

\usepackage{upgreek}
\usepackage[version=3]{mhchem} 
\usepackage[T1]{fontenc}       
\usepackage{color}
\usepackage{multirow}
\usepackage{makecell}
\usepackage{amssymb}
\usepackage{amsmath}
\usepackage[normalem]{ulem}
\graphicspath{{figures/}}

\SectionNumbersOn

\author{Mathieu Luisier}
\email{*mluisier@iis.ee.ethz.ch}
\author{Cedric Klinkert}
\author{Sara Fiore}
\author{Jonathan Backman}
\author{Youseung Lee}
\author{Christian Stieger}
\author{\'Aron Szab\'o}
\affiliation{Integrated System Laboratory, ETH Zurich,
  CH-8092 Zurich, Switzerland}

\title{Field-Effect Transistors based on 2-D Materials: a Modeling Perspective}

\begin{document}

\section{Introduction}\label{sec:intro}

\subsection{Future of Moore's Law}

In 1965, Gordon Moore, one of Intel's co-founders, formulated in
\cite{moore} his now famous law that states that the number of
transistors per integrated circuit (IC) doubles every 18 to 24
months. This miracle has been made possible over the last 50+ years
thanks to an aggressive
scaling of the dimensions of silicon-based metal-oxide-semiconductor
field-effect transistors (MOSFETs), as reviewed in \cite{moore_review}.
Till 2003, this miniaturization followed Dennar's scaling law
(\cite{dennard}), which consisted in reducing the spatial dimensions
(width, gate length, oxide thickness) and power supply of every new
MOSFET generation by 30\%. In other words, these quantities were
multiplied by a factor of 0.7$\times$ from one generation to the
other. As a consequence, the power density of ICs stayed constant over
the years, while their performance kept increasing, driven by the
shortening of the transistor gate length. Dennard's scaling stopped
however in 2003 at the so-called 130 nm technology node (TN) because
the supply voltage of transistors could no more be reduced at the same
pace as their dimensions. The sub-threshold slope ($SS$), which
indicates how rapidly the electrical current of logic switches can be
increased between their OFF and ON states explains this phenomenon. In
conventional MOSFETs, it is limited to 60 mV/dec at room temperature:
the gate voltage must be swept by at least 60 mV to vary the current
by one order of magnitude. 

The impossibility to push $SS$ below this limit in MOSFETs
forced the semiconductor industry to maintain relatively large supply
voltages (above 1 V), thus leading to significant increases of the
power and heat dissipation of electronic devices, see \cite{pop}. At the
circuit level, the end of Dennard's law could be partly compensated by
decreasing the clock frequency of ICs to reduce the power dissipation
and by combining multiple cores together with shared 
memory to augment the computational capabilities. This was the
beginning of the ``multicore crisis'', an on-going era with
energy-efficient, parallel, but sequentially slower multicore computers
than at the beginning of the 2000's. At the device level, from 2003
onwards, it was observed that ``simply'' scaling the size of
transistors was no more sufficient to enhance their operation, in
particular their switching speed. Since then, different technology
boosters have therefore been introduced to ensure that the performance
improvements historically brought by Moore's scaling law could
continue, as summarized in \cite{kuhn}
\begin{itemize}
\item Strain engineering as in \cite{strain}: from the 90 nm TN,
  strain has been used in Si MOSFETs to alter the bandstructure of
  electrons and holes, with an increase of their channel mobility as a
  result; 
\item High-$\kappa$ oxide layers as in \cite{oxide}: at the 45 nm TN,
  SiO$_2$, the native oxide of Si, was replaced by high-$\kappa$
  dielectric layers such as HfO$_2$ that can provide larger gate
  capacitances together with lower leakage currents;
\item 3-D FinFETs as in \cite{finfet}: till the 32 nm TN,
  transistors were 2-D planar structures with a single-gate
  contact. In 2011, they became 3-D FinFETs with a triple-gate
  configuration, offering a higher immunity against short-channel
  effects, in particular source-to-drain tunneling. 
\end{itemize}
The benefits of multicore architectures, strain, high-$\kappa$ dielectrics,
and FinFETs are best visible in the electronic products that we use on
a daily basis, be they cell phones, tablets, or laptops.
Quantitatively, these benefits can be assessed by considering the
Top500 list of supercomputers available in \cite{top500}. This list
ranks all participating machines based on their performance when
running the LINPACK benchmark, see \cite{linpack}. What is measured is
the number of floating point operations (Flop) that are processed per
second (Flop/s) to solve a linear system of equations on the hardware
of interest. Back in 1993 (first Top 500 list), the largest
supercomputer in the world achieved 60$\times$10$^9$ Flop/s, i.e. 60
GFlop/s. Today's cell phones, with a peak performance larger than
10$^{12}$ Flop/s (1 TFlop/s), are about 15 to 20 times more powerful
than this machine. For its part, the current largest supercomputer in
the world, as of November 2020, reaches 440$\times$10$^{15}$ Flop/s
(440 PFlop/s), according to \cite{top500}. 

\begin{figure}[ptbh]
\centering
\includegraphics[width=\linewidth]{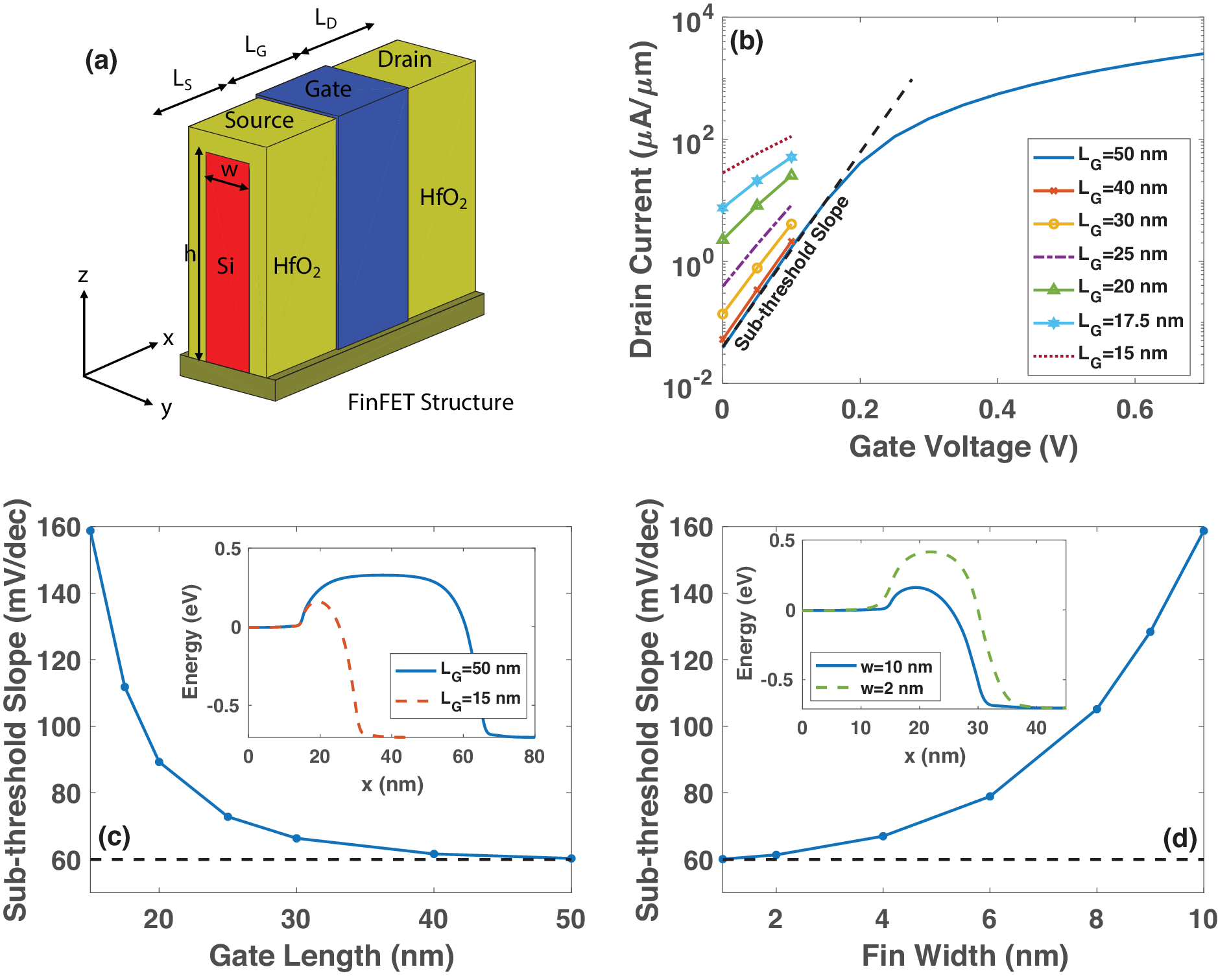}
\caption{(a) Schematic view of an $n$-type 3-D FinFET with a fin
  height $h$=40 nm and width $w$ comprised between 1 and 10 nm. The
  $L_S$=15 nm, $L_G$, and $L_D$=15 nm quantities refer to the length
  of the source extension, gate region, and drain extension,
  respectively. The source and drain are doped with a donor
  concentration $N_D$=10$^{20}$ cm$^{-3}$. The Si channel (red) is aligned
  with the $<$110$>$ crystal axis and is surrounded by HfO$_2$ layers
  of thickness $t_{ox}$=4 nm. (b) Transfer characteristics $I_D$-$V_{GS}$
  at $V_{DS}$=0.7 V of the FinFET in (a) with $w$=10 nm and $L_G$=50
  nm (blue line). The current was normalized by 2$\times$h+w. The
  currents at low $V_{GS}$ are also given for FinFETs with 15$\leq
  L_G\leq$40 nm. The sub-threshold slope $SS$ is indicated by the
  dashed black line. (c) Sub-threshold slope of the FinFET in (a) as a
  function of $L_G$ for $w$=10 nm. Inset: electrostatic potential
  energy at $L_G$=15 and 50 nm. (d) Same as (c), but as a function
  of $w$ for $L_G$=15 nm. Inset: electrostatic potential energy at
  $w$=10 and 2 nm.}
\label{fig:finfet}
\end{figure}

While Si FinFETs have established themselves as the transistors of
reference since 2011, they might lose this position in the future,
when their gate length will shrink below 15 nm. One of the main
challenges associated with the scaling of such devices is illustrated
in Fig.~\ref{fig:finfet} using a simplified device structure with realistic
dimensions, as schematized in sub-plot (a). At a gate length $L_G$=50
nm, a fin width of 10 nm together with a fin height of 40 nm yield a
subthreshold slope $SS$=60.8 mV/dec at room temperature, close to the
theoretical limit (sub-plot (b)). If the cross section dimensions are
kept the same, but the gate length reduced from 50 to 15 nm, $SS$ rapidly
explodes, reaching 160 mv/dec for the shortest considered $L_G$, as can
be seen in sub-plot (c). The observed deterioration is caused by a
loss of electrostatic control, which can be related to a parameter called
``natural channel length'' $\lambda_N$ and defined in
\cite{appenzeller} as
\begin{equation}
  \lambda_N=\sqrt{\frac{\epsilon_{Si}}{N\epsilon_{ox}}T_{Si}T_{ox}},
\end{equation}
where $\epsilon_{Si}$, $\epsilon_{ox}$, $N$, $T_{Si}$, and $T_{ox}$
are the relative permittivity of Si, that of the oxide layer, the number
of gate contacts, the thickness of the Si channel, and that of the
oxide surrounding it, respectively. Short-channel effects can be
prevented if the gate length $L_G$ is at least 6$\times$ larger than
$\lambda_N$. Hence, $\lambda_N$ should be made as small as possible to
conveniently scale transistors.

In FinFETs, the number of gate contacts, $N$, is equal to 3. If Si
remains the channel material and HfO$_2$ the dielectric of choice,
only $T_{Si}$ and $T_{ox}$ can be made thinner. Because decreasing
$T_{ox}$ would ultimately lead to high OFF-state gate leakage
currents, the only viable solution to scale the $L_G$ of FinFETs
appears to be a reduction of 
$T_{Si}$, which corresponds to the fin width $w$. This is what has
been done in Fig.~\ref{fig:finfet}(d). When pushing $w$ down to 1 nm,
a $SS$ value of $\simeq$60 mV/dec can be achieved, which would readily
allow to push the gate length of FinFETs (well) below 15 nm, while maintaining a
good electrostatics integrity. However, reliably fabricating fin
structures with a width of 1 nm only is extremely demanding. Surface
roughness, single impurities, or interface traps are all expected to
play a non-negligible role at this scale and to negatively impact the
Si channel mobility, see \cite{takagi}. Besides these effects,
device-to-device variability might become a real issue for
ultra-scaled FinFET sizes.

\begin{figure}[ptbh]
\centering
\includegraphics[width=\linewidth]{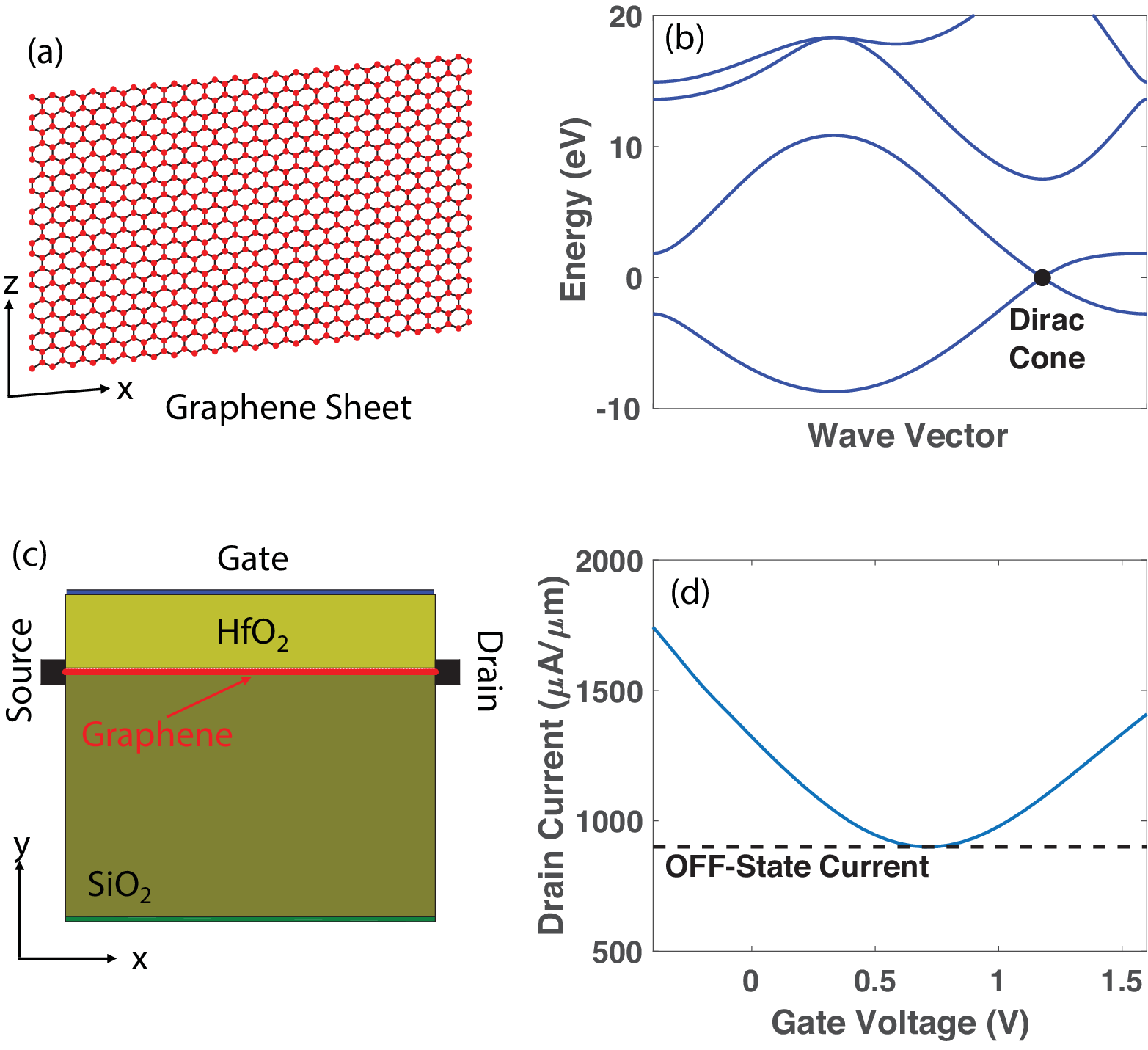}
\caption{(a) 2-D graphene flake. The red dots represent atoms, the
  black lines atomic bonds. (b) Bandstructure of graphene along the
  $M-\Gamma-K-M$ line of the Brillouin Zone. The location of the Dirac
  cone is indicated by a black dot. (c) Schematic view of a single-gate graphene
  field-effect transistor (GFET) with a gate length $L_G$=15
  nm. Schottky contacts are assumed for the source and drain with a
  barrier height of 0.25 eV. The red dots refer to carbon atoms. The
  channel is separated from the gate contact by a HfO$_2$ layer of
  thickness $t_{ox}$=3 nm and is deposited on a SiO$_2$ substrate. (d)
  Transfer characteristics $I_D$-$V_{GS}$ of the GFET in (c) at
  $V_{DS}$=0.2 V. The OFF-state current cannot be brought below 900
  $\mu$A/$\mu$m, a much too high value.}
\label{fig:graphene}
\end{figure}

\subsection{The Potential of 2-D Materials}

Instead of thinning the width of semiconductors that normally have a
3-D unit cell, 2-D materials with a naturally flat atomic structure
might be more promising to create tomorrow's transistors with a gate
length of 15 nm and below (\cite{fiori}). Such compounds are
characterized by an almost perfect electrostatic control, no surface
roughness, and no dangling bonds. Graphene, a carbon monolayer
with a honeycomb lattice, is an excellent example of a 2-D material. Its
existence was confirmed experimentally in 2004 through mechanical
exfoliation, as the dependence of its conductance on an external
electric field, see \cite{novoselov}. This finding motivated the
fabrication of graphene field-effect transistors (GFETs), as in \cite{lemme},
but the absence of a band gap in graphene does not enable to
fully switch these devices off, as indicated in Fig.~\ref{fig:graphene}. A
dispersionless Dirac cone where the conduction and valence bands of
touch each other can be noticed in the bandstructure plot, at
the $K$-point of the Brillouin Zone. Because of it, the current flow
from source to drain can never be completely blocked by the
gate-modulated potential barrier, as in Si or III-V MOSFETs.

\begin{figure}[ptbh]
\centering
\includegraphics[width=\linewidth]{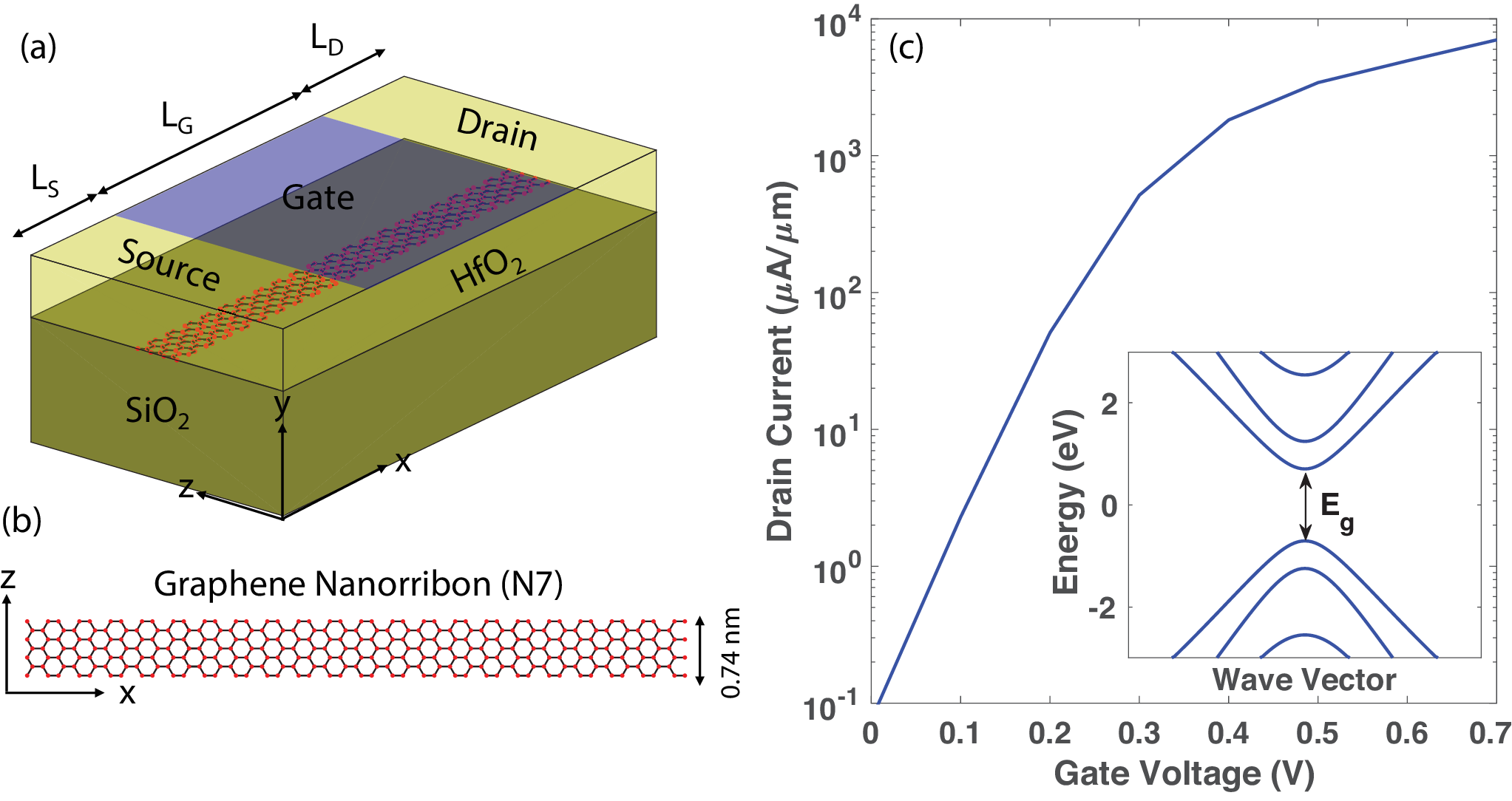}
\caption{(a) Schematic view of a graphene nanoribbon field-effect
  transistor (GNRFET). It is identical to the structure in
  Fig.~\ref{fig:graphene}, except that the 2-D graphene flake was
  replaced by a N7 nanoribbon of width $w$=0.74 nm and that the source
  and drain extensions of length $L_S$=$L_D$=25 nm are doped with a
  donor concentration $N_D\approx$10$^{13}$ cm$^{-2}$. (b) Atomic structure
  of the N7 graphene nanoribbon in (a). (c) Transfer characteristics
  $I_D$-$V_{GS}$ at $V_{DS}$=0.7 V of the GNRFET in (a). The OFF-state
  current is fixed to $I_{OFF}$=0.1 $\mu$A/$\mu$m (normalized by the
  nanoribbon width $w$=0.74 nm). The inset shows the bandstructure of
  the N7 nanoribbon whose band gap $E_g$=1.408 eV.}
\label{fig:gnr}
\end{figure}

A band gap can be opened up in graphene if it is patterned into a
quasi 1-D nanoribbon, either lithographically as in \cite{kim} or
chemically as in \cite{empa}. While the former approach is more straightforward to
realize, it tends to produce graphene nanoribbons (GNRs) wider than
20 nm whose transport properties suffer from detrimental line edge
roughness. The width of chemically-derived GNRs can be reduced below 1
nm with an excellent control of their edges and reproducible
electrical characteristics. Transistors made of such 1-D
nanostructures can exhibit ON/OFF current ratios greater than 10$^{6}$ as
well as ON-state currents in the order of 2000 $\mu$A/$\mu$m
(\cite{dai}). The simulation of a N7 GNR field-effect transistor (GNRFET)
with a width $w$=0.74 nm is shown in Fig.~\ref{fig:gnr} as an example.
The investigated logic switch has a gate length $L_G$=15 nm. It
provides a steep sub-threshold slope $SS$=67.5 mV/dec, a high ON-current
$I_{ON}$=6.87 mA/$\mu$m at an OFF-current $I_{ON}$=74 nA/$\mu$m
and supply voltage $V_{DD}$=0.7 V. To deliver such a performance, the
source and drain extensions of the GNRFET were doped with a donor
concentration of $N_D\approx$10$^{13}$ cm$^{-2}$, which might not be
attainable experimentally. Furthermore, it is rather difficult to
obtain low contact resistances in GNRFETs and the mass production of
ultra-narrow structures is very tedious. Needed are materials that
present themselves in the form of relatively easily manufacturable
large-scale flakes, as graphene, but that display a band gap
compatible with logic applications.

Monolayers of transition metal dichalcogenides (TMD) of $MX_2$
composition, where $M$ is a transition metal and $X$ a chalcogene,
fulfill these conditions with their thickness below 1 nm, band gap
between 1 and 2 eV, high carrier mobility, and availability as large 
flakes (\cite{akinwande}). They are therefore often seen as serious
contenders to continue Moore's scaling law in the ``more-than-Moore''
category (\cite{schwierz}). In TMDs, each layer is composed of $M$ atoms
surrounded by two $X$ atoms. The inter-atomic bonds within each layer
are covalent, whereas van der Waals forces maintain adjacent layers
together in few-layer structures. TMDs can adopt different crystal
lattices (and symmetry groups), from the usually semiconducting 2H
(hexagonal) phase to the typically metallic 1T (trigonal) or 1T'
(modified trigonal) phase, going through other phases such as 3R
(rhombohedral) or 2M (monoclinic). One lattice is generally more
stable than the others, but phase transition can be triggered by 
external fields, strain, or doping, see \cite{phase}.

Among all existing TMD materials and configurations, single-layer
MoS$_2$ was exfoliated for the first time in 1986 with a scotch tape
(\cite{joensen}), but it was only after the first experimental
demonstration in 2011 of a properly working 2H single-gate monolayer
MoS$_2$ transistor with $SS$=74 mV/dec and an ON/OFF current ratio
larger than 10$^{8}$ in \cite{kis} that TMD started to receive a wide
attention from the scientific community. Since then, transistors with
a WSe$_2$ (\cite{javey}), WS$_2$ (\cite{allain}), MoTe$_2$
(\cite{seabaugh}), MoSe$_2$ (\cite{wang}), ReS$_2$ (\cite{liu}), HfSe$_2$,
or ZrSe$_2$ (\cite{mleczko}) channel have been reported as well, to cite
few examples. TMD monolayers were used in most cases, except for the
MoTe$_2$ (down to 6L), HfSe$_2$, and ZrSe$_2$ (down to 3L) ones, which
relied on few-layer structures. Among these 2-D semiconductors, some
are better suited to obtain $n$-type transistors, e.g. MoS$_2$,
ReS$_2$, HfSe$_2$, or ZrSe$_2$, others lend themselves more naturally
to $p$-type devices (WSe$_2$ and MoTe$_2$), whereas MoSe$_2$ is rather
ambipolar once crystal defects have been repaired. The type of each
TMD is primarily determined by two effects: the metallic contacts
attached to it, which affect the Schottky barrier height at the
metal-semiconductor interfaces, and the dielectric environment around
it, which can transfer electrons or holes to the channel.

The complementary logic at the core of electronic circuits requires
both $n$- and $p$-type transistors. Hence, the polarity of 2-D TMDs
should be modifiable, which can be done through doping. For
instance, a potassium-doped WSe$_2$ channel becomes $n$-type. It can
then be combined with a $p$-type WSe$_2$ transistor to give rise to a
fully 2-D inverter as in \cite{javey2}. Doping TMDs and more generally 2-D
materials remains however a difficult task. Several options exist for
that. A back-gate can be inserted to modulate the band edges of the
source and drain extensions with respect to the contact Fermi
levels (electrostatically-induced doping). Alternatively, ion
implantation (\cite{suh}), ion intercalation (\cite{kappera}), or charge
transfer from, e.g. ionic liquid (\cite{allain_2}) can be utilized.
Nevertheless, none of this approach currently allows to reach doping
concentrations as high as in Si.

Apart from doping, TMDs face several other challenges that prevent
device engineers from accessing their intrinsic performance and
compare it to that of Si FinFETs. One of them is the obtention of high
quality, large-scale monolayers. Already in the 1960's, mechanical
exfoliation via scotch tape was applied to isolate layered materials
(\cite{frindt}). Only small flakes can be gained with this rather 
straightforward technique, thus hindering mass production of TMD
transistors. Lately, impressive progresses have been made with the
help of the chemical vapor deposition (CVD) (\cite{najmaei}) and
metal-organic CVD (\cite{kang}), which has been shown to produce large
monolayer areas with high carrier mobilities. High-quality TMD flakes
in the $\mu$m$^2$ range can also be generated with atomic layer
deposition (ALD) (\cite{bol}).

Other challenges related to the fabrication of high-performance,
TMD-based transistors will be discussed in Section~\ref{sec:chall}. As
addressing most of them will necessitate significant technology
improvements, device simulation can be used in the meantime to
provide invaluable insights into the physics of these logic
switches, predict their performance limit under ideal conditions, and
support the on-going experimental activity. This Chapter will first
introduce a suitable modeling approach for that in Section
\ref{sec:mod} before presenting results of recently undertaken
theoretical investigations in Section \ref{sec:perf}. The 
latter are not restricted to TMDs, but will also explore novel 2-D
semiconductors that could become the channel material of future,
ultra-scaled $n$- and $p$-type transistors.

\begin{figure}[ptbh]
\centering
\includegraphics[width=\linewidth]{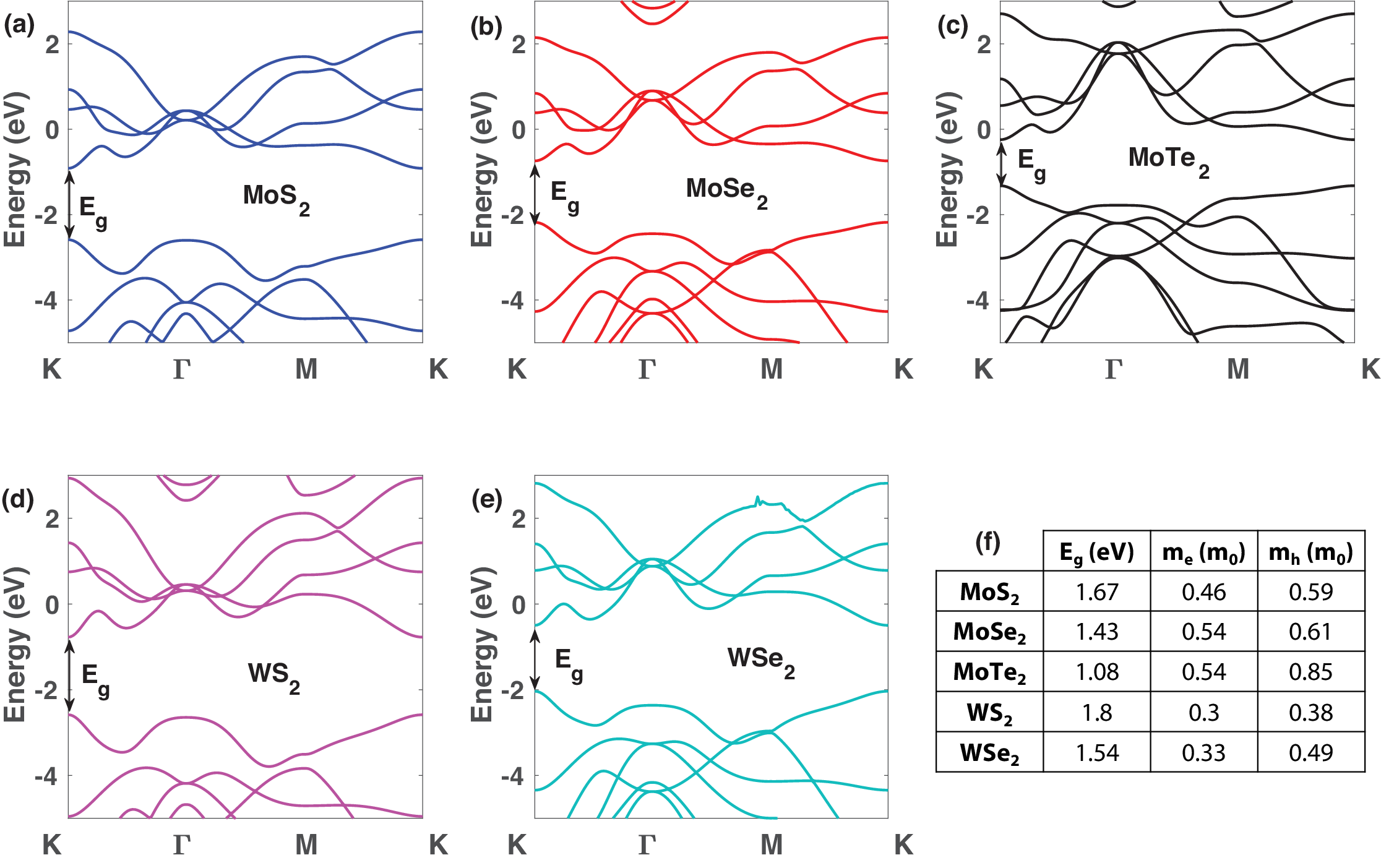}
\caption{(a) Bandstructure of MoS$_2$ as computed with
  density-functional theory (DFT). (b) Same as (a), but for
  MoSe$_2$. (c) Same as (a), but for MoTe$_2$. (d) Same as (a), but
  for WS$_2$. (e) Same as (a), but for WSe$_2$. (f) Table summarizing
  the electron and hole effective masses of the conduction and valence
  band extrema as well as the band gap of the 2-D materials in (a) to (e).}
\label{fig:tmd_bs}
\end{figure}

\section{Modeling Approach}\label{sec:mod}

\subsection{Requirements and State-of-the-Art}

A critical ingredient of device simulation, regardless of the chosen
approach, is the bandstructure of the different materials that
constitute the domain of interest. The bandstructure enters either
directly or indirectly into the physical model, via the Hamiltonian
matrix $H$ or the effective mass $m^*$ of the different components,
respectively. These quantities are connected through Schr\"odinger's
equation:
\begin{equation}
  H(k)\cdot\psi_k(r)=E(k)\psi_k(r),
\label{eq:schroed}
\end{equation}
\begin{equation}
  \frac{1}{m^*}=\left.\frac{1}{\hbar^2}\frac{d^2E(k)}{dk^2}\right|_{dE/dk=0},
\end{equation}
where $k$ refers to the electron wave vector, $\psi_k(r)$ to the wave
function of the system at position $r$ and wave vector $k$, $E(k)$ to
the corresponding $k$-dependent band dispersion, and $\hbar$ to Planck's
reduced constant. The effective masses are extracted at band extrema
that are characterized by the condition $dE/dk=0$. In
Figs.~\ref{fig:finfet}, \ref{fig:graphene}, and \ref{fig:gnr}, the
Hamiltonian matrices of the investigated FinFET, GFET, and GNRFET were
constructed in the effective mass approximation (EMA) for Si and in the
single-$p_z$ orbital scheme of \cite{wallace} for graphene. Both models
are computationally very attractive, pretty accurate for the systems 
mentioned above, but unfortunately not ideal for most 2-D materials,
as will be explained in the following paragraphs.

The bandstructure of selected TMDs is shown in Fig.~\ref{fig:tmd_bs},
together with their band gap and effective masses. All these
quantities were calculated with density-functional theory (DFT), as
proposed by \cite{dft}, an \textit{ab initio} (from first-principles)
method that does not require any input parameters, except for the
initial atomic unit cell (AUC) of the material under
consideration. This provided AUC is first
relaxed so that all atoms occupy stable positions. The corresponding
electron density is then self-consistently computed with Poisson's
equation. Finally, all electronic bands are extracted from the
obtained DFT Hamiltonian $H_{DFT}$. Although very accurate, DFT still
relies on several approximations, among them the exchange-correlation
functional, here the PBE one of \cite{pbe}.

What can be clearly seen in Fig.~\ref{fig:tmd_bs} is that the
bandstructures of TMDs exhibit complex features such as multiple
valleys separated by a small energy interval, strongly non-parabolic
bands, and in some cases band anisotropy (the effective masses depend
on the crystal orientation). To capture all these
effects, a quantum mechanical simulation approach is absolutely
necessary. In other words, a Hamiltonian matrix as in
Eq.~(\ref{eq:schroed}) must be assembled to describe the electronic
properties of the desired device. On their side,
neither classical nor semi-classical methods can shed light 
on the physics of ultra-thin 2-D materials where electrons and holes
are confined over narrow dimensions with dimensions below 0.5
nm. Despite these shortcomings, drift-diffusion calculations can
fairly well reproduce experimental data for large-scale TMD flakes, if
the available material parameters are properly calibrated (\cite{grasser}). 

The first quantum mechanical study of a monolayer MoS$_2$ transistor
was reported in \cite{salahuddin}: it was shown that this material
could outperform Si as logic switch with a sub-15 nm gate length. To
come to this conclusion a quantum transport simulator based on
the EMA and the Non-equilibrium Green's Function (NEGF) formalism
proposed by \cite{datta} was employed. The functionality of this
simulator is similar to the one that
produced the data in Fig.~\ref{fig:finfet}. We would like to emphasize
that NEGF is one of the most popular and powerful techniques to examine
charge transport in nanoscale devices. The basic NEGF equations will
be reviewed in Section \ref{sec:negf}. At about the same time, a
simplified solver based on the straightforward top-of-the-barrier
model confirmed these results and revealed that other TMDs might be
competitive as well (\cite{guo}). However, because EMA assumes parabolic
bands, the satellite conduction band valleys of TMDs could not be
properly accounted for in these works. Furthermore, the atomic
granularity of single-layer crystals was totally ignored, EMA being
continuous.

Device simulators implementing a semi-empirical full-band model
represents the next level of accuracy and a significant improvement
over the EMA as they include more than a single parabolic
band. Hence, the k$\cdot$p as in \cite{kormanyos} or tight-binding (TB)
method as in \cite{hguo} or \cite{klimeck} can be used to construct the Hamiltonian
matrix of TMDs. As each atom or discretization point is described by a
set of $N_{orb}$ orbitals, the simulation time is in the order of
$N_{orb}^3$ times longer than with EMA. Such an increase is still
acceptable from a computational point of view because the number of
neighbors per point, which determines the band width of the
Hamiltonian matrix and the size of the numerical blocks to manipulate,
usually remains small. On the negative side, both k$\cdot$p and
tight-binding models must be first parameterized, a sometimes tedious
operation that is not always successful. Moreover, getting physically
meaningful parameters to connect two different materials, e.g. a TMD
and the metallic contact attached to it, can be rather
complicated. Finally, k$\cdot$p in its usual forms (4$\times$4 to
8$\times$8) tends to be restricted to one symmetry point in the
Brillouin Zone and does not capture the atomic nature of TMDs.

Going up the ladder, a DFT solver relying on a localized basis set can
be coupled to a quantum transport simulator based on the NEGF
formalism (\cite{qw0}). DFT+NEGF ticks all accuracy boxes (atomistic, 
full-band, \textit{ab initio}, capable of treating interfaces and
defects, $\cdots$), but the computational burden of such an approach
explodes for large structures composed of several hundreds of
atoms. On the one hand, more orbitals per atom are needed. On the
other hand, the inter-atomic interactions extends over much longer
distances than in tight-binding or k$\cdot$p. In summary, neither
$H_{EMA}$, $H_{k\cdot p}$, $H_{TB}$, nor $H_{DFT}$ Hamiltonian
matrices are optimal to simulate the electrical behavior of
transistors with a 2-D channel material.

\begin{figure}[ptbh]
\centering
\includegraphics[width=\linewidth]{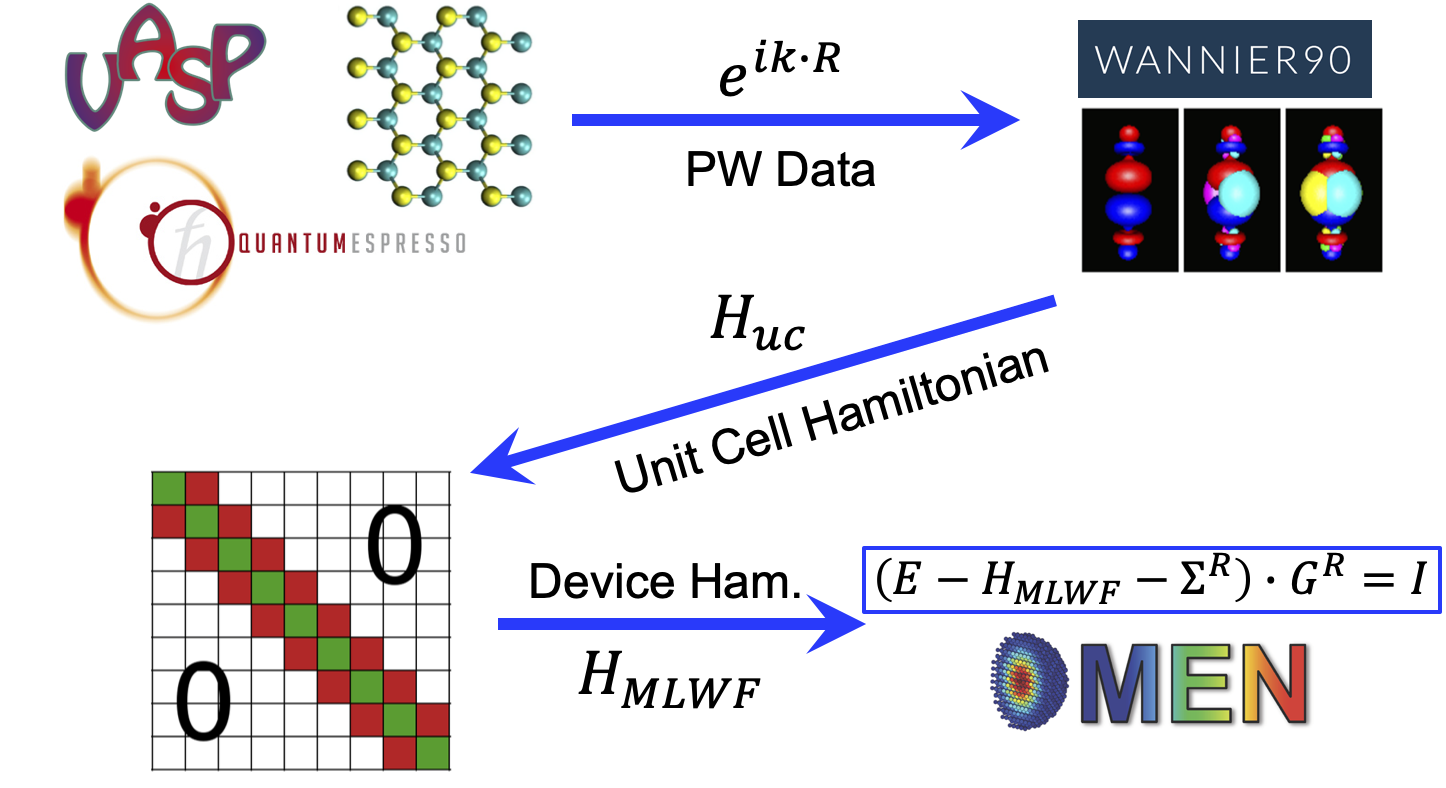}
\caption{\textit{Ab initio} scheme to simulate transistors with a 2-D
  channel material. A DFT calculation of a representative 2-D unit
  cell is first performed with a plane-wave (PW) code such as VASP (\cite{vasp})
  or Quantum ESPRESSO (\cite{qe}). The results are then converted to a
  maximally localized Wannier function basis with the wannier90 tool
  (\cite{wannier}), which applies a unitary transformation. Next, the
  produced unit cell Hamiltonian, $H_{uc}$, is scaled up to the device
  dimensions, as in \cite{stieger}. Finally, the obtained $H_{MLWF}$ is
  passed to a quantum transport simulator, e.g. OMEN (\cite{omen}).}
\label{fig:upscale}
\end{figure}

\subsection{Maximally Localized Wannier Functions (MLWFs)}

A model that has the same accuracy as DFT and a computational
complexity comparable to tight-binding or k$\cdot$p would be ideal to
probe 2-D materials as next-generation logic switches. Maximally
localized Wannier functions (MLWFs), as introduced by \cite{mlwf},
satisfy both conditions, as was first demonstrated in \cite{register} 
for various TMDs. Since this early work, MLWFs have become very
popular among the 2-D research community, as in
\cite{bruzzone,szabo2,pizzi,pourtois,kubis,yoon,dong,afzalian}, to
mention a few relevant examples. The MLWF method is as accurate as
DFT plane-wave (PW) calculations from which they are derived, they require
a small number of basis elements per atom, and the distance over which
they decay is very short. The numerical treatment of the resulting
Hamiltonian matrices, $H_{MLWF}$, is therefore facilitated. Thanks to
these unique features, MLWFs allow to simulate large transistor
structures made of thousands of atoms within reasonable computational
times. They can be seen as a first step towards fully \textit{ab
  initio} device investigations.

\begin{figure}[ptbh]
\centering
\includegraphics[width=\linewidth]{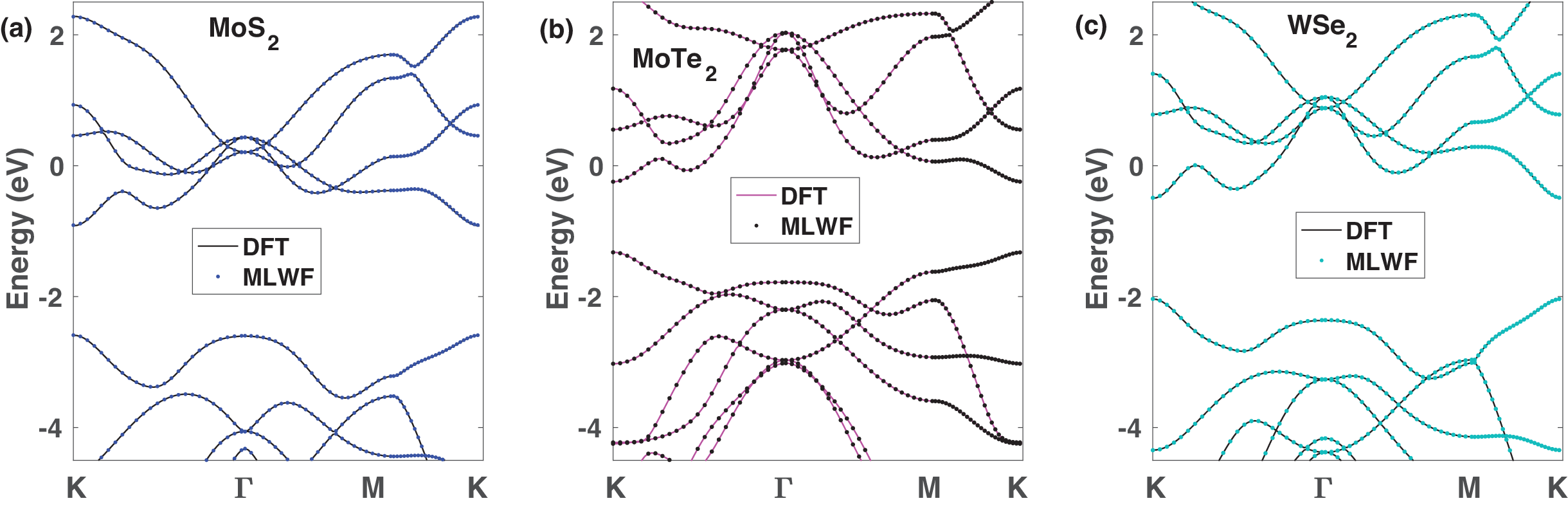}
\caption{(a) Bandstructure of MoS$_2$ as computed with
  DFT using the VASP tool (solid black lines) and after a
  transformation of the plane-wave results into a set of maximally
  localized Wannier functions (MLWFs, blue dots). (b) Same as (a), but
  for a MoTe$_2$ monolayer. (c) Same as (a) and (b), but for WSe$_2$.}
\label{fig:mlwf_1}
\end{figure}

The principle of MLWF-based quantum transport simulations is
summarized in Fig.~\ref{fig:upscale}. The whole process starts with a
plane-wave DFT calculation of a primitive unit cell that best
represents the geometry of the targeted atomic system. Different tools
such as VASP (\cite{vasp}) or Quantum ESPRESSO (\cite{qe}) can be used for
that purpose. The produced eigenenergies and eigenvectors are then
transformed into a set of MLWFs with the wannier90 package
(\cite{wannier}). The required unitary transformation is exact so that
the bandstructure obtained in the PW and MLWF basis sets are theoretically
identical, the only difference being that MLWFs only return a sub-set
of all PW bands, in the present case those required to evaluate
transport properties. Practically, small discrepancies might emerge
due to the truncation of long-ranging interactions. They have a
limited impact on the results. As illustrations, the bandstructures of
MoS$_2$, MoTe$_2$, and WSe$_2$ are plotted in Fig.~\ref{fig:mlwf_1},
as computed with DFT and after a transformation into MLWFs. Excellent
agreement between both data sets can be observed.

This PW-to-MLWF conversion gives rise to small Hamiltonian blocks that
describe the coupling of the chosen unit cell with itself and with its
neighboring cells. Those blocks must be upscaled to form a block
tri-diagonal Hamiltonian matrix corresponding to the device to be
simulated. Such an upscaling scheme is described in \cite{stieger}. The
obtained $H_{MLWF}$ is perfectly suitable for quantum transport
simulations, its increased band width being partly compensated by the
fact that less orbitals per atom are required. As a consequence, the
same numerical algorithms as with tight-binding can still be employed
to solve the NEGF equations. A quantum transport solver such as OMEN
(see \cite{omen}) can do that as it has been specifically designed to handle
large-scale nanostructures from first-principles (\cite{sc15}). Note
that the procedure outlined in Fig.~\ref{fig:upscale} works for any
exchange-correlation functional, e.g. the generalized gradient
approximation with PBE parameterization (\textit{GGA-PBE}) of \cite{pbe},
hybrid functionals (\textit{HSE06}) of \cite{hybrid}, or the \textit{GW}
plus Bethe-Salpeter equation (GW-BSE) of \cite{gw}. The initial DFT run
will be longer with \textit{HSE06} or \textit{GW}, but the time for
the transport calculation is not affected by this choice.

\subsection{Towards \textit{Ab Initio} Quantum Transport Simulations}\label{sec:negf}

In the previous Section, the construction of a MLWF-based Hamiltonian
matrix was presented. As last step, $H_{MLWF}$ should be passed to a
quantum transport (QT) solver to perform \textit{ab initio} device
simulations. Various QT methods have been proved effective, among them
the solution of the Wigner Transport Equation (\cite{wigner}), Pauli's
Master Equation (\cite{pauli}), the Quantum Transmitting Boundary Method
(QTBM) (\cite{qtbm}), and, of course, the Non-equilibrium Green's
Function (NEGF) formalism (\cite{datta}) that was already mentioned
above. As all results in Section \ref{sec:perf} have been obtained
with NEGF, the equations governing this transport approach will now be
briefly introduced, with emphasis on 2-D material applications. 

To describe electron transport within the NEGF framework, the following
non-linear system of equations must be solved
\begin{eqnarray}
\left\{
\begin{array}{l}
\left(E-H_{MLWF}(k_z)-\Sigma^{RB}(E,k_z)-\Sigma^{RS}(E,k_z)\right)\cdot G^{R}(E,k_z)=I,\\
G^{\gtrless}(E,k_z)=G^{R}(E,k_z)\cdot\left(\Sigma^{\gtrless B}(E,k_z)+\Sigma^{\gtrless S}(E,k_z)\right)\cdot G^{A}(E,k_z).
\end{array}
\right.
\label{eq:negf}
\end{eqnarray}
In Eq.~(\ref{eq:negf}), the $G$'s represent the electron Green's
Functions. They depend on the electron energy $E$ and momentum $k_z$
as well as on the MLWF Hamiltonian matrix $H_{MLWF}$. The $k_z$ momentum
models the flake direction that is orthogonal to the transport axis
and that is assumed periodic. The $G$'s can be of four
different types, retarded ($R$), advanced ($A$), lesser ($<$), or
greater ($>$). The $\Sigma$'s refer to the corresponding self-energies
where the superscript $B$ stands for boundary and $S$ for
scattering. With $\Sigma^B$, the coupling of the simulation domain with
contact electrodes is captured. This self-energy can be computed with
so-called decimation techniques (\cite{sancho}) or more advanced
schemes, for example through eigenvalue problems (\cite{luisier}) or
contour integrals, as in \cite{beyn}. Its scattering counterpart $\Sigma^S$
can include different interaction mechanisms such as electron-phonon,
impurity, or interface roughness scattering, see \cite{lake}. 

\begin{figure}[ptbh]
\centering
\includegraphics[width=\linewidth]{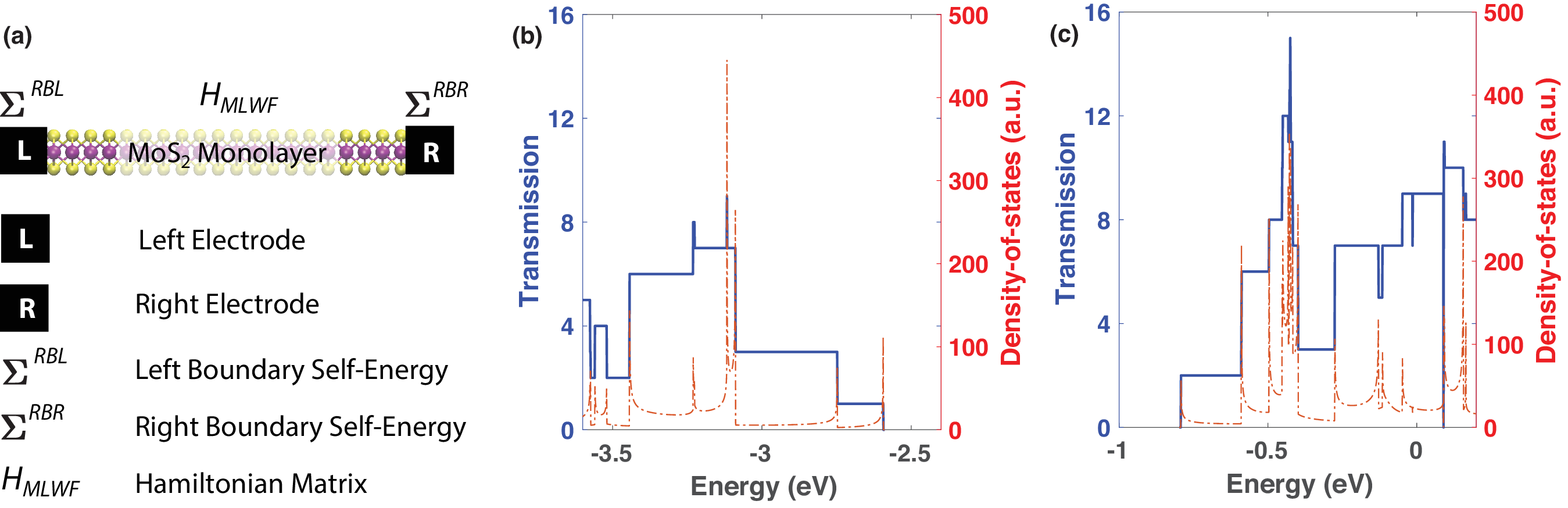}
\caption{(a) Schematic view of a MoS$_2$ monolayer described by a
  Hamiltonian matrix $H_{MLWF}$. The plot shows how contacts are
  accounted for through boundary self-energies ($\Sigma$
  matrices). (b) Transmission function (blue line) and
  density-of-states (dashed red line) of the valence band of a MoS$_2$
  monolayer under flat band conditions at $k_z$=0. These results were
  obtained with the computational scheme of Fig.~\ref{fig:upscale}(c)
  Same as (b), but for the conduction band of a MoS$_2$ monolayer.}
\label{fig:trans}
\end{figure}

Equation (\ref{eq:negf}) can be efficiently solved with a recursive
algorithm that constructs the Green's Functions from one side of the
device to the other in two steps, see \cite{rgf}. In case of ballistic
transport, i.e. in the absence of interactions with other carriers,
impurities, rough surfaces or crystal vibrations,
only the retarded Green's function $G^R$ needs to be calculated. With its
knowledge, both the density-of-states (DOS) and transmission function
($TE$) of the considered system can be evaluated, from which the carrier
density and the electronic current can be derived as in \cite{datta}. The
DOS and $TE$ of a MoS$_2$ monolayer are provided in Fig.~\ref{fig:trans}
under flat band conditions, evaluated at a single momentum point,
$k_z$=0, for both the conduction and valence bands. The role of
$H_{MLWF}$ and $\Sigma^{RBL}$/$\Sigma^{RBR}$, the boundary
self-energies, is highlighted as well. As expected under such 
circumstances, the transmission function displays a step-like behavior 
and effectively counts the number of bands available in the left and
right contacts. Each time a transmission channel turns on, the DOS
peaks, followed by an exponential decay that is only interrupted
by the next peak.

If scattering should be accounted for, the lesser and
greater Green's Functions $G^{\gtrless}$ in Eq.~(\ref{eq:negf}) must also
be computed. These quantities must be solved self-consistently with
the scattering self-energies $\Sigma^{\gtrless S}$ as they depend on
each other. This has been done for example in \cite{szabo1} for
single-, double-, and triple-layer MoS$_2$ where electron-phonon
scattering was treated at the \textit{ab initio} level. The same
approach was extended in \cite{stieger2} to model self-heating effects
and the formation of local hot spots in various field-effect
transistors with a TMD monolayer as channel material.

All results presented this Chapter have been obtained in the presence of
electron-phonon scattering, but the chosen model relies on a
simplified and phenomenological approach with one single phonon energy
$\hbar\omega$=40 meV and a scattering strength $D_{e-ph}$ comprised
between 25 and 125 meV (\cite{klinkert}):
\begin{equation}
\Sigma^{\lessgtr S}(k_z,E)=D^2_{e-ph}\left(n_{\omega}G^{\lessgtr}(k_z,E+\hbar\omega)+(n_{\omega}+1)G^{\lessgtr}(k_z,E-\hbar\omega)\right). 
\label{eq:sigma}
\end{equation}
In Eq.~(\ref{eq:sigma}) $n_{\omega}$ is the phonon's Bose-Einstein
distribution function. A dissipative scattering mechanism is needed
to avoid a negative differential resistance (NDR) behavior in the
$I_D$-$V_{DS}$ output characteristics of the 2-D FETs, which has never
been experimentally observed at room temperature. NDR originates from
the bandstructure of 2-D materials, which often exhibits several
narrow bands that cannot propagate if the electrostatic potential
undergoes large variations from source to drain (\cite{gnani}). It is an
artifact of the ballistic approximation. The inclusion of
electron-phonon scattering helps get rid of this non-physical effect
by connecting bands that would otherwise be independent from
each other (\cite{szabo1}). Accounting for the ``real'' electron-phonon
interactions would be more accurate, but gathering the required phonon 
bandstructures and coupling elements is computationally very
demanding, as the self-consistent calculation of the scattering
self-energies. For all these reasons, the model of
Eq.~(\ref{eq:sigma}) was adopted.

Finally, it should be noted that all simulations were performed at
room temperature with the metal gate work function adjusted so that the
OFF-state current is fixed to 0.1 $\mu$A/$\mu$m. Perfectly ohmic
contacts are assumed (no resistance), except if mentioned otherwise.

\begin{figure}[ptbh]
\centering
\includegraphics[width=\linewidth]{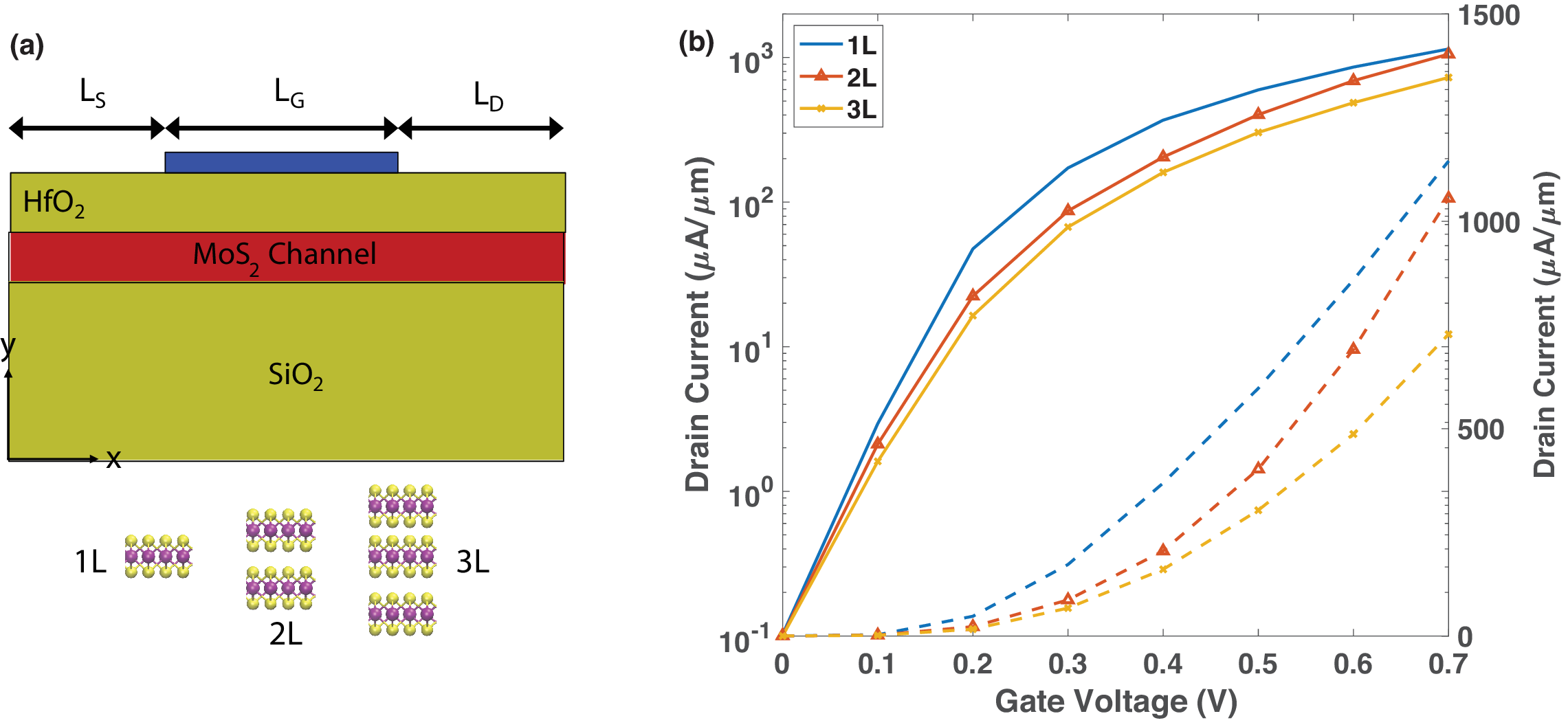}
\caption{(a) Schematic view of a $n$-type single-gate transistor with a
  single- (1L), bi- (2L), and triple- (3L) layer of MoS$_2$ as
  channel. The gate length $L_G$ measures 15 nm, the source ($L_S$)
  and drain ($L_D$) extensions 10 nm with a donor concentration
  $N_D$=5$\times$10$^{13}$ cm$^{-2}$. (b) Transfer characteristics $I_D$-$V_{GS}$ at
  $V_{DS}$=0.7 V of the MoS$_2$ transistors in (a), on a logarithmic
  (solid lines) and linear (dashed lines) scale.}
\label{fig:mos2}
\end{figure}

\section{2-D Device Performance Analysis}\label{sec:perf}

\subsection{MoS$_2$ and other TMDs}

The first 2-D material under the \textit{ab initio} microscope is
MoS$_2$, the TMD whose monolayer form was initially shown to provide 
excellent transistor characteristics in \cite{kis}. Before assessing the
performance of MoS$_2$ transistors with respect to other 2-D
materials, we would like to underline the importance of the channel
thickness. In Fig.~\ref{fig:mos2}, a simplified device structure is
schematized. Its channel is either made of a MoS$_2$ monolayer,
bilayer, or trilayer, whose thickness is approximately 0.6, 1.2, and
1.8 nm, respectively. The corresponding transfer characteristics are
shown on the right side of the plot. A MLWF Hamiltonian matrix was
constructed for each channel configuration. The \textit{ab initio}
simulations reveal that the monolayer structure has the highest
ON-state current (1.15 mA/$\mu$m), followed by the bilayer (1.06
mA/$\mu$m), and finally the trilayer one (0.73 mA/$\mu$m). Normally, the
opposite order would be expected as the transport effective mass
$m_{trans}$ decreases as the number of stacked layers increases.
However, the gate contact loses part of its control efficiency at
larger channel thicknesses, as already demonstrated in
Fig.~\ref{fig:finfet}.

This deterioration is best reflected in the $SS$ value of each
transistor, which goes from 68.6 mV/dec in the monolayer case to 75.6
mV/dec in the bilayer and 82.8 mV/dec in the trilayer. The gate
contact can very well modulate the height of the potential barrier in
the layer that is the closest to it, but its influence decreases as
carriers are situated away from it. Hence, the benefit of smaller
effective masses in few-layer structures is washed out by the poorer
electrostatics of thicker channels. If we compare these results to
those of Fig.~\ref{fig:finfet}(d), we notice a substantially larger
$SS$ than in the FinFET with a 1-nm wide fin, even for the monolayer FET. The
presence of a triple-gate in the FinFET case explains the better
scalability of this device. Adding a second gate to the 2-D MoS$_2$
FETs produces the same effect and leads to an almost perfect
electrostatic control in mono-, bi- and tri-layer structures down to a
gate length $L_G$=10 nm (\cite{szabo1}). Note that in this publication,
different DFT models were used than here.

\begin{figure}[ptbh]
\centering
\includegraphics[width=\linewidth]{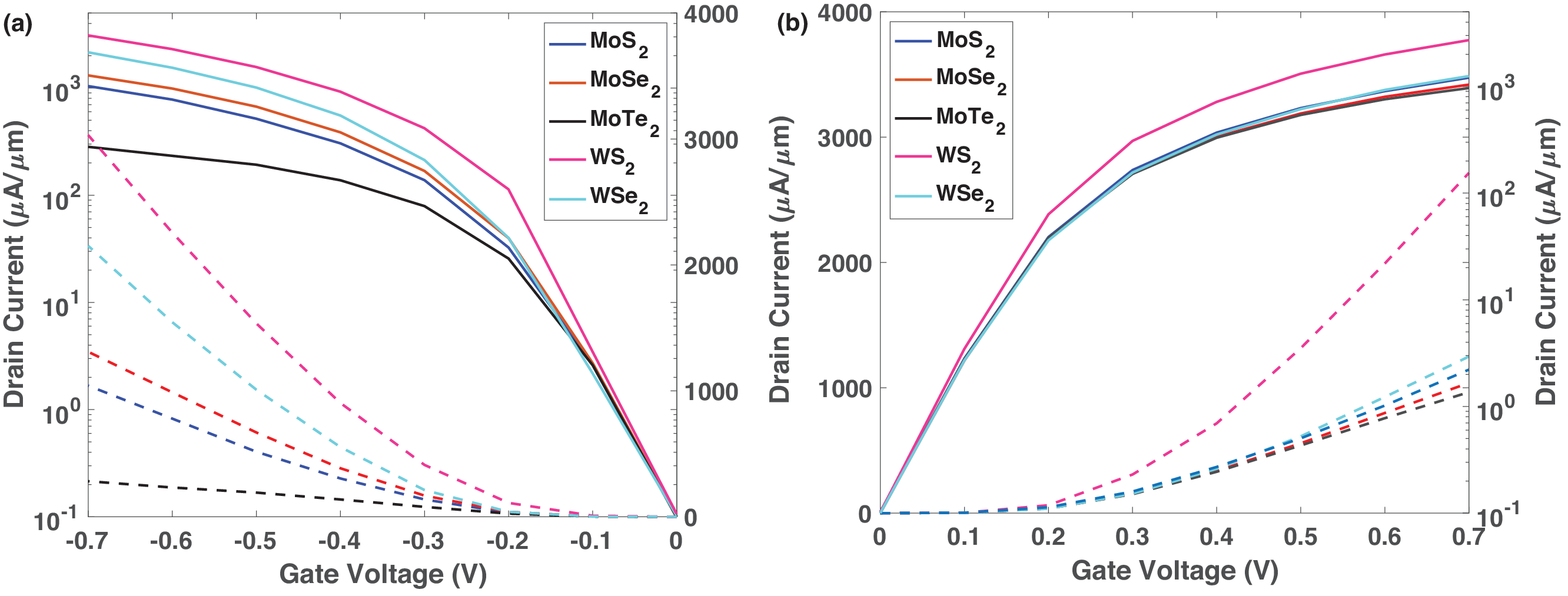}
\caption{(left) Transfer characteristics $I_D$-$V_{GS}$ at
  $V_{DS}$=-0.7 V of $p$-type monolayer TMD field-effect transistors
  (FETs) with a gate length $L_G$=15 nm, both on a linear (dashed
  lines) and logarithmic (solid lines) scale. The following TMD are
  represented: MoS$_2$ (blue), MoSe$_2$ (red), MoTe$_2$ (black),
  WS$_2$ (magenta), and WSe$_2$ (cyan). (right) Same as on the left,
  but for $n$-type monolayer TMD FETs at $V_{DS}$=0.7 V.}
\label{fig:TMD_I_V}
\end{figure}

MoS$_2$ has become the most popular TMD, as confirmed by the number of
publications dedicated to it, but it is not necessarily the most
promising one. Experimentally, no ON-state current larger than
700 $\mu$A/$\mu$m has ever been reported for monolayer MoS$_2$
\cite{pop2}. This value was obtained at $V_{DS}$= 5 V, $V_{GS}$=30 V,
and in a device with a gate length $L_G$=380 nm and an ON/OFF current
ratio larger than 10$^6$. Other TMDs have therefore also been
investigated. The proposed \textit{ab initio} simulation approach can
also be applied to them, as shown in Fig.~\ref{fig:mlwf_1}. The MLWF
results perfectly reproduces the plane-wave DFT calculations for all
TMDs. Taking advantage of that, the transfer characteristics 
$I_d$-$V_{gs}$ at $V_{ds}$=0.7 V of $n$- and $p$-type MoSe$_2$,
MoTe$_2$, WS$_2$, and WSe$_2$ field-effect transistors were computed
as well. They are displayed in Fig.~\ref{fig:TMD_I_V} and compared to
those of MoS$_2$. All devices have a structure similar to the one in
Fig.~\ref{fig:mos2}(a) with a gate length $L_G$=15 nm and a TMD
monolayer as channel material.

First, it can be seen that almost all 2-D FETs have a sub-threshold
slope in the order of 70 mV/dec, despite the relatively short $L_G$
and the presence of a single gate contact. There is one exception,
WS$_2$, whose $SS$ is smaller ($\sim$65 mV/dec). This is not a
consequence of a better electrostatic control, but of the influence of
narrow energy bands. As explained above, they can lead to the presence
of NDR as well as to too low $SS$ values, even below 60
mV/dec. A physical parameter called ``pass factor'' allows to quantify
the importance of these bands, as explained in \cite{klinkert}.
The electron-phonon scattering model of Eq.~(\ref{eq:sigma})
is expected to eliminate these artefacts, but for some
2-D materials, e.g. WS$_2$, it does not fully succeed. Increasing the
electron-phonon coupling can improve the situation. However, at the
same time, this affects the ON-state current, preventing a fair
comparison with other 2-D materials.

The ON-state current values are more broadly distributed than the
sub-threshold slopes. Generally, it can be observed that the W-based
TMDs perform better than the Mo-based ones, due to lower effective masses and
therefore faster carriers, as illustrated in
Fig.~\ref{fig:tmd_bs}. This trend is not altered if electron-phonon
scattering is included because W atoms have a larger mass than
Mo's so that their oscillation amplitude is smaller, as their
probability to interact with free carriers (\cite{iedm16}). Another
\textit{ab initio} theoretical study came to a different conclusion,
predicting that the mobility of WS$_2$ would be significantly impacted
by electron-phonon scattering (\cite{sohier}). The small energy difference
between the conduction band minimum of this material and its satellite
valleys is the reason behind this discrepancy. As this energy
separation strongly depends on the choice of the DFT functional,
different calculations might have different outcomes.  

\begin{table}
  \centering
  \begin{tabular}{c|c|c|c|c|c|}
    &MoS$_2$&MoSe$_2$&MoTe$_2$&WS$_2$&WSe$_2$\\
    \hline
    $n$-$I_{ON}$ (mA/$\mu$A)&1.15&1.04&1.11&2.71&1.25\\
    \hline
    $n$-$SS$ (mV/dec)&68.6&70.1&71&65.3&69.9\\
    \hline
    $p$-$I_{ON}$ (mA/$\mu$A)&1.04&1.31&0.28&3.03&2.15\\
    \hline
    $p$-$SS$ (mV/dec)&70.6&70&70.4&66&75.2\\
    \hline
  \end{tabular}
\caption{Summary of the $n$- and $p$-type ON-state currents $I_{ON}$
  and of the sub-threshold slopes $SS$ extracted from the
  $I_D$-$V_{GS}$ transfer characteristics in Fig.~\ref{fig:TMD_I_V}.}
\label{tab:1}  
\end{table}

The $I_{ON}$ and $SS$ of all simulated 2-D TMD FETs are summarized in
Table \ref{tab:1}. Overall, under ideal conditions, only one 2-D TMD
offers an ON-state current significantly larger than 1 mA/$\mu$m,
both in its $n$- and $p$-type configuration, WS$_2$: the $n$-type
$I_{ON}$ is 2.71 mA/$\mu$m, the $p$-type one 3.03 mA/$\mu$m. Due to
the aforementioned issues with the sub-threshold region of these FETs
and the possibility that the WS$_2$ mobility might be lower than
expected, it must be concluded that 2-D TMDs are probably not the best
candidates to replace Si as future, more-than-Moore
transistors. Further challenges, but also opportunities to improve
their properties will be discussed in Section \ref{sec:chall}.

\begin{figure}[ptbh]
\centering
\includegraphics[width=\linewidth]{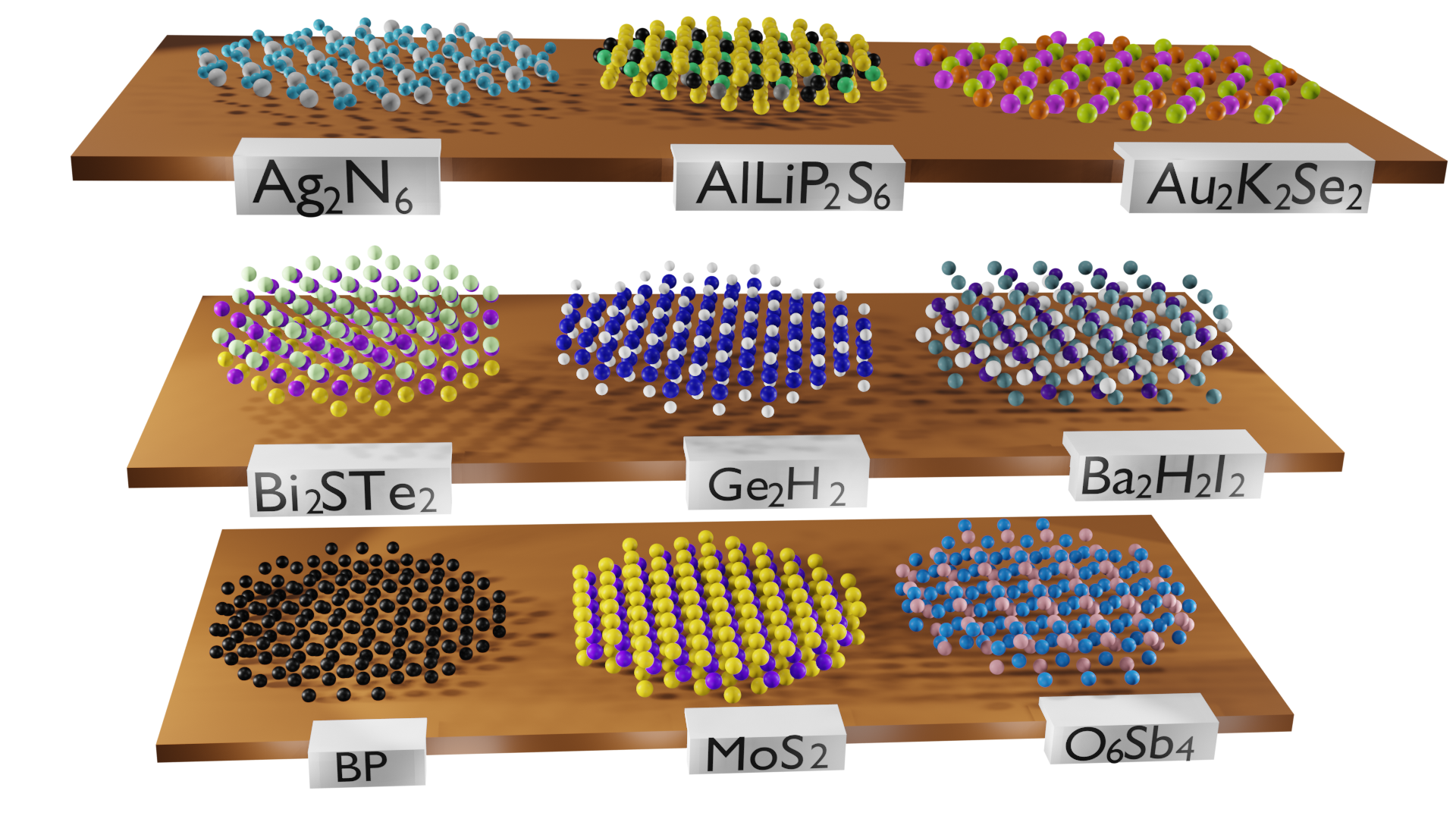}
\caption{Atomic unit cells of few selected 2-D materials
  from \cite{mounet}.}
\label{fig:shelf}
\end{figure}

\subsection{Novel 2-D Materials}

Besides TMDs, other 2-D materials suitable for logic applications have
emerged over the years, starting with black phosphorus, a monolayer of
phosphorus (BP) atoms with a buckled honeycomb lattice (\cite{ye}). Its
high hole mobility makes it particularly attractive as $p$-type
field-effect transistor, as demonstrated experimentally in
\cite{zhang}. While BP has established itself as the most prominent
alternative to TMDs, other novel 2-D materials have earned themselves
a place under the spotlight, for example silicene in \cite{silicene},
germanene in \cite{germanene}, antimonene in \cite{antimonene}, InSe
in \cite{inse}, or Bi$_2$O$_2$Se in \cite{bi2o2se}. Many others are
also in the pipeline, see \cite{franklin,2d_review}.

On the theoretical side, a high throughput (HT) investigation by
\cite{mounet} revealed that more than 1,800 2-D materials might exist,
among them about 1,000 easily exfoliable monolayers. To come up with
these numbers the authors considered a large set of 3-D parent
crystals from the Inorganic Crystal Structure Database (ICSD,
\cite{icsd}) and the Crystallographic Open Database (COD, \cite{cod}).
They then applied geometrical criteria to identify layered compounds
and extract 2-D children from them, tested the stability of the latter 
in vacuum by computing their phonon bandstructure, and finally
classified them according to their inter-layer binding energy $E_b$ and
the influence of van der Waals forces. Those of interest have a low
$E_b$ and their layers are kept together by non-covalent, van der
Waals bonds. All notable 2-D materials (graphene, TMDs, BP, $\cdots$)
were recovered by the HT study of \cite{mounet}. They are accompanied
by components with a huge variety of band gaps (from metals to
oxides), effective masses (from isotropic to strongly anisotropic
bandstructures), and monolayer thicknesses (from one to several
repeatable atomic layers). Few examples are proposed in
Fig.~\ref{fig:shelf}.

\begin{figure}[ptbh]
\centering
\includegraphics[width=\linewidth]{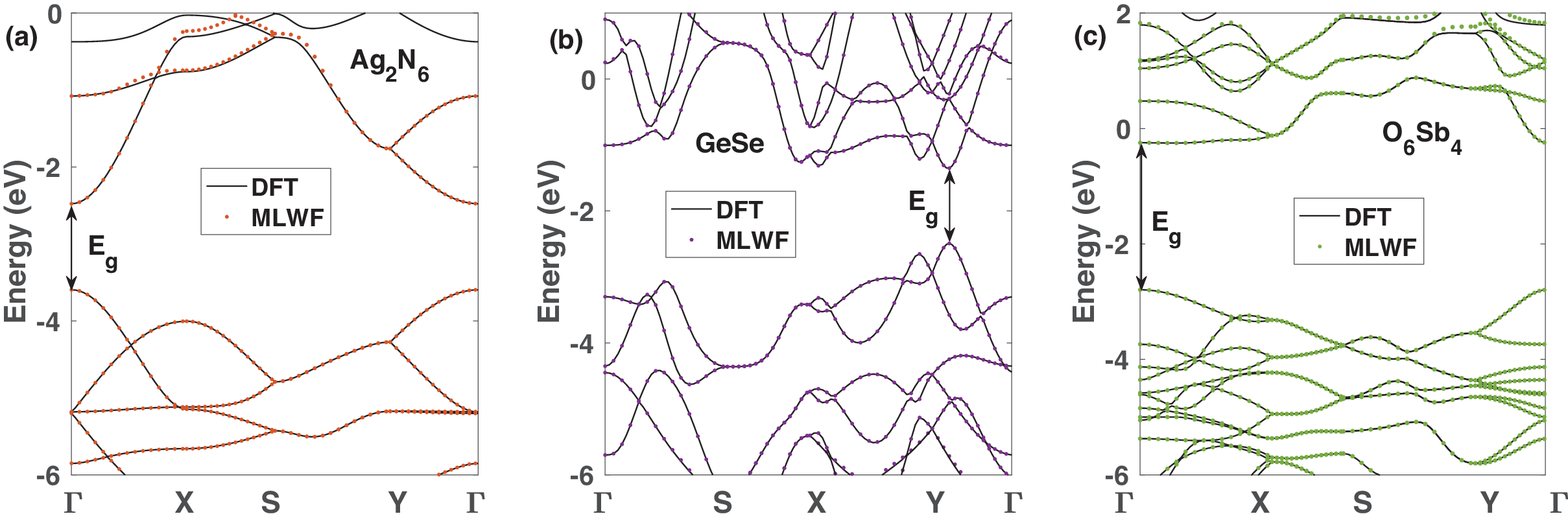}
\caption{(a) Bandstructure of Ag$_2$N$_6$ as computed with
  DFT using VASP (solid black lines) and after a transformation of the
  plane-wave results into MLWFs (orange dots). (b) Same as (a), but
  for a GeSe monolayer. (c) Same as (a) and (b), but for O$_6$Sb$_4$.}
\label{fig:mlwf_2}
\end{figure}

Independently from this study, the potential of 2-D materials beyond
TMDs and BP as field-effect transistors has been evaluated through
device simulations, with empirical and \textit{ab initio} models. The
related literature is abundant and not all important contributions can
be listed here. Nevertheless, the following examples are deemed
representative: Bi$_2$O$_2$Se in \cite{quhe}, monochalcogenides in
\cite{lake2}, group IV in \cite{wvdb}, group V in \cite{pizzi}, as
well as on more exotic 2-D materials such as Tl$_2$O in \cite{heine}.
To be able to compare the performance of transistors made of these very
different 2-D components, a uniform modeling approach would be
preferable, for instance the one described in Section 
\ref{sec:mod}. It is validated in Fig.~\ref{fig:mlwf_2} for
non-conventional 2-D monolayers, Ag$_2$N$_6$, GeSe, and
O$_6$Sb$_4$. As for TMDs, the DFT and MLWF bandstructure results agree
very well so that everything is in place to conduct a large-scale and
systematic performance comparison of 2-D FETs. These results were
originally published in \cite{klinkert}. All approximations,
simulation results, and extracted material parameters are
presented in its Supplementary Information.

\begin{figure}[ptbh]
\centering
\includegraphics[width=0.7\linewidth]{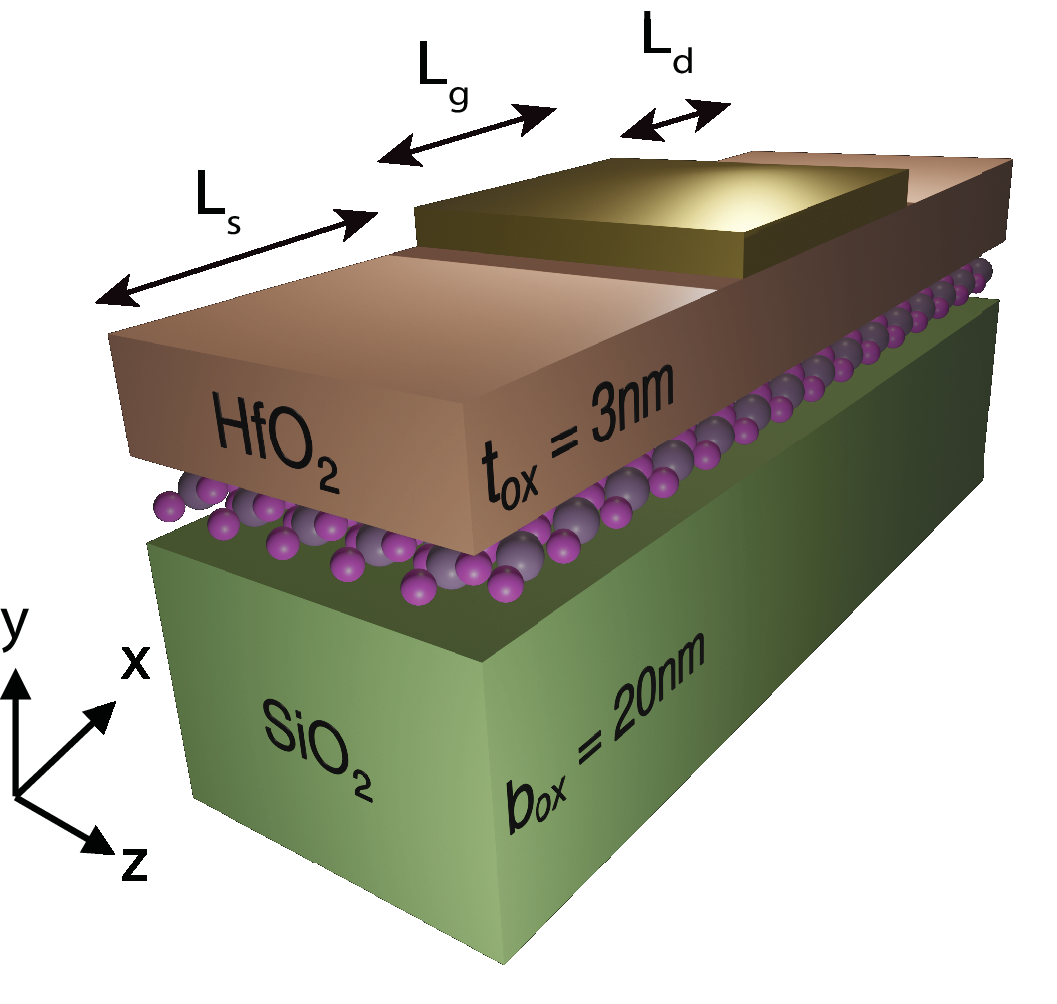}
\caption{Schematic view of the single-gate 2-D monolayer field-effect
  transistors investigated in this Chapter. The total device length 
  is set to 40 nm, with a gate length $L_G$ varying between 15 and 5
  nm. The source and drain extensions are doped with a donor $N_D$ or
  acceptor $N_A$ concentration of 5$\times$10$^{13}$ cm$^{-2}$ for the
  $n$- and $p$-type configuration, respectively. A $t_{ox}$=3 nm thick
  high-$\kappa$ dielectric layer (HfO$_2$) separates the 2-D channel,
  which is deposited on a SiO$_2$ box with $t_{box}$=20 nm, from the
  gate contact. Transport occurs along the $x$-axis, $y$ is a
  direction of confinement, whereas $z$ is assumed to be periodic and
  gives rise to the $k_z$-dependence in Eq.~(\ref{eq:negf}). Adapted
  from \cite{klinkert}.}
\label{fig:mosfet}
\end{figure}

In this context, we designed a realistic 2-D transistor structure, as
in Fig.~\ref{fig:mosfet}, and defined a set of targeted figures of
merits (FOM). The dimensions and specifications of this single-gate
FET derive inspiration from the International Roadmap for Devices and 
Systems (IRDS) in \cite{irds} for the year 2025, i.e. a gate length
$L_G$=15 nm, a supply voltage $V_{DD}$=0.7 V, and an equivalent oxide
thickness (EOT) of 0.6 nm. This EOT is achieved through a 3 nm HfO$_2$
dielectric layer with a relative permittivity $\epsilon_R$=20. The 2-D
materials are deposited onto a SiO$_2$ substrate with a thickness
$t_{box}$=20 nm and $\epsilon_R$=3.9. To ensure a satisfactory
electrostatic control, the source and drain extensions of the FETs are
doped with a donor/acceptor concentration $N_{D/A}$=5$\times$10$^{13}$
cm$^{-2}$. Such high values cannot be achieved experimentally for the
moment, see \cite{pop2} or \cite{suh2}, but could be in the future,
for example by combining different doping techniques. The SiO$_2$ and
HfO$_2$ domains do not enter the NEGF equations, they are treated as
perfectly insulating layers that only impact the electric field profile. 

In terms of FOM, we are looking for 2-D materials offering an ON-state
current larger than 3 mA/$\mu$m, both in their $n$- and $p$-type
configurations, at an OFF-state current $I_{OFF}$=0.1 $\mu$A/$\mu$m. At
the same time, the sub-threshold slope $SS$ should not exceed 80
mV/dec at a gate length $L_G$=15 nm. We selected 100 different 2-D
materials from the database of \cite{mounet} that could potentially
reach these objectives. As first criterion, we singled out thin 
semiconductor monolayers with a small number of atoms in their
primitive unit cell (PUC). No compounds thicker that 1.5 nm and with
more than 30 atoms in their PUC were considered. Secondly, using
the bandstructure calculation results of \cite{mounet}, we further
restricted ourselves to 2-D materials with a band gap larger than 1 eV
and, if possible, anisotropic conduction band minima and/or valence band
maxima so that a low transport and high density-of-states effective
mass are simultaneously obtained. Finally, only monolayers that
are stable in vacuum were retained, i.e. those whose phonon
bandstructure does not have negative branches around the
$\Gamma$-point.

\begin{figure}[ptbh]
\centering
\includegraphics[width=\linewidth]{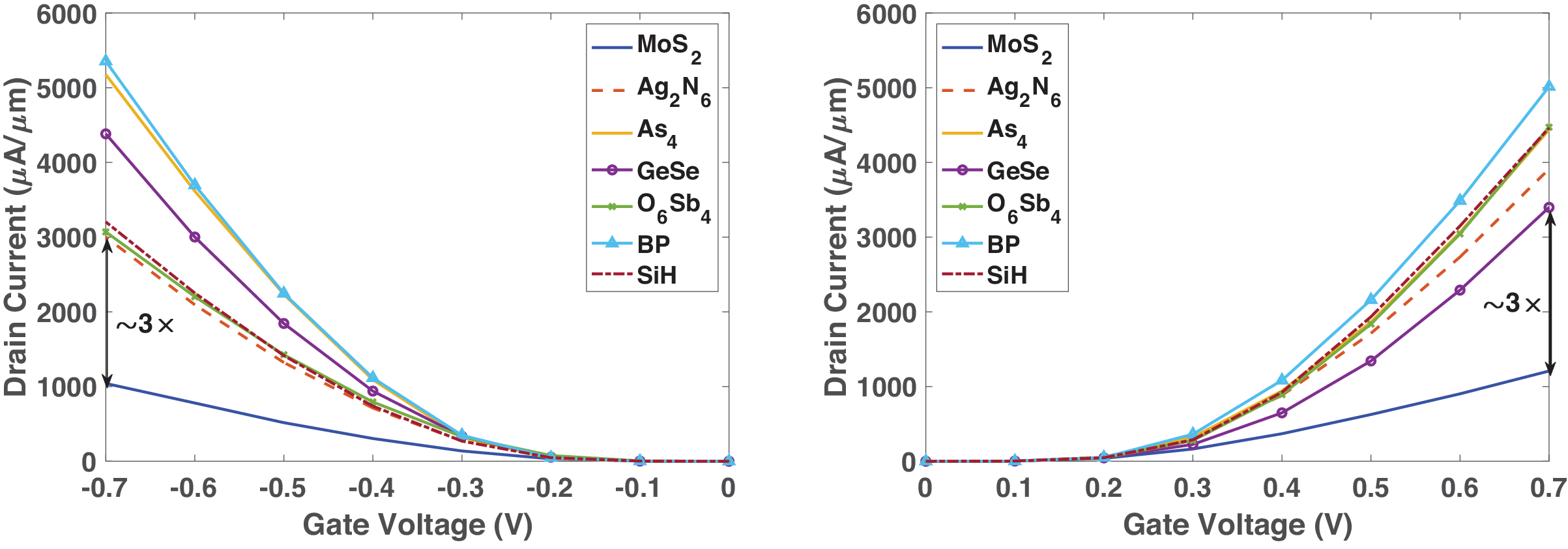}
\caption{(left) Transfer characteristics $I_D$-$V_{GS}$ at
  $V_{DS}$=-0.7 V of $p$-type monolayer FETs with a gate length
  $L_G$=15 nm. They are made of promising 2-D materials (Ag$_2$N$_6$,
  As$_2$, GeSe, O$_6$Sb$_4$, black phosphorus, SiH) and compared to
  a MoS$_2$ device. (right) Same as on the left, but for $n$-type
  monolayer FETs at $V_{DS}$=0.7 V. For both transistor
  configurations, the ON-state current of the promising 2-D materials
  increases by a factor of $\sim$3$\times$ (at the same $I_{OFF}$=0.1
  $\mu$A/$\mu$m), as compared to MoS$_2$.}
\label{fig:2D_I_V}
\end{figure}

Fig.~\ref{fig:2D_I_V} reports the $n$- and $p$-type transfer
characteristics ($I_D$-$V_{GS}$) of 6 promising 2-D materials that
satisfy the conditions listed above: Ag$_2$N$_6$, As$_2$, GeSe,
O$_6$Sb$_4$, black phosphorus (BP), and SiH (silicane). Under ideal
conditions, they all deliver $I_{ON}\geq$3 mA/$\mu$m, which is about
3$\times$ larger than MoS$_2$, BP even reaching ON-state currents in the
order of 5 mA/$\mu$m. The $SS$ of all transistors is
around 70 mV/dec at $L_G$=15 nm, which is 10 mV/dec lower than the
target that was set. In total, out of the 100 examined 2-D materials,
13 arrive at the desired level of performance. Their FOM and effective
masses are summarized in Table \ref{tab:2}. None of the conventional
TMDs from Fig.~\ref{fig:tmd_bs} (MoS$_2$, MoSe$_2$, MoTe$_2$, WS$_2$,
and WSe$_2$) belongs to that group, although WS$_2$ gets very close to
it ($n$-type $I_{ON}$=2.71 mA/$\mu$m and $p$-type $I_{ON}$=3.03
mA/$\mu$m), as can be seen in Table \ref{tab:1}. Two less common TMDs,
HfS$_2$ and ZrS$_2$, seem to have a higher potential, their $n$- and
$p$-type ON-state currents being above 3 mA/$\mu$m in the ballistic
limit of transport.

\begin{figure}[ptbh]
\centering
\includegraphics[width=\linewidth]{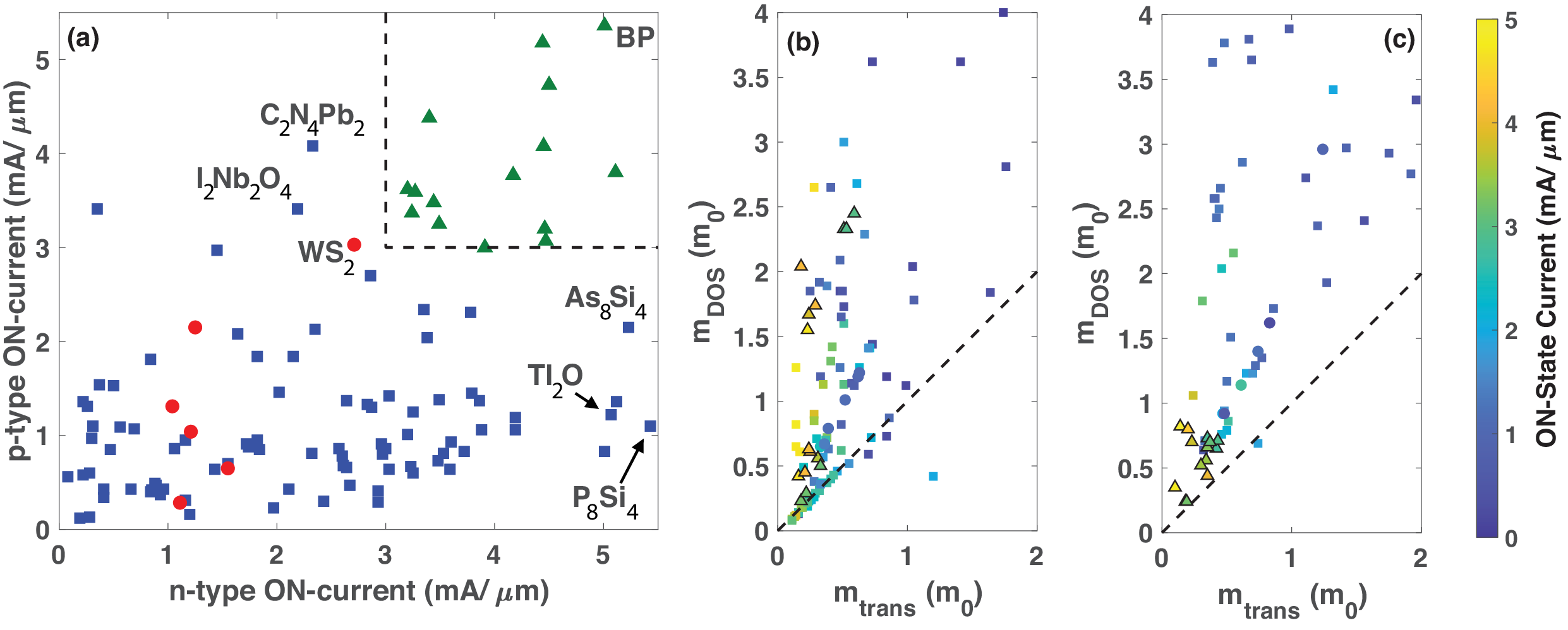}
\caption{(a) ``$n$-type ON-state current vs. $p$-type ON-state
  current'' for the 100 selected 2-D materials. The red circles refer
  to the usual TMDs, while the green triangles encompass the
  best-performing compounds, i.e. those with an $n$- and $p$-type
  $I_{ON}$ larger than 3 mA/$\mu$m. (b) ON-current of the 100
  simulated $n$-type 2-D FETs in (a) as a function of their
  transport $m_{trans}$ and density-of-states $m_{DOS}$ effective
  masses. The same marker shapes as in (a) are used. The current
  magnitude is encoded in the marker color (see color bar on the
  right). The black dashed line indicates 2-D materials with an
  isotropic bandstructure ($m_{trans}$=$m_{DOS}$). (c) Same as (b),
  but for $p$-type 2-D FETs. Adapted from \cite{klinkert}.}
\label{fig:ion}
\end{figure}

To give an overview of all 2-D materials that have been simulated, we
put together their ``$n$-type $I_{ON}$ vs. $p$-type $I_{ON}$''
characteristics in Fig.~\ref{fig:ion}(a). This plot allows to
rapidly identify the 13 best-performing candidates as they are
situated in the upper right corner delimited by black dashed
lines. Overall, several 2-D materials (39) exhibit an 
ON-state current larger than 3 mA/$\mu$m in their $n$-type form,
e.g. P$_8$Si$_4$ (5.43 mA/$\mu$m), As$_8$Si$_4$ (5.23 mA/$\mu$m), or
Tl$_2$O (5.09 mA/$\mu$m), much less (17) as $p$-FET,
e.g. C$_2$N$_4$Pb$_2$ (4.08 mA/$\mu$m) or I$_2$Nb$_2$O$_4$ (3.41
mA/$\mu$m). As for the Si-based CMOS technology, the fact that more
2-D compounds have a high $n$-type rather than $p$-type ON-state
current indicates that fabricating high-performance pFETs might be a
challenging task in the flat land as well. 

\begin{figure}[ptbh]
\centering
\includegraphics[width=\linewidth]{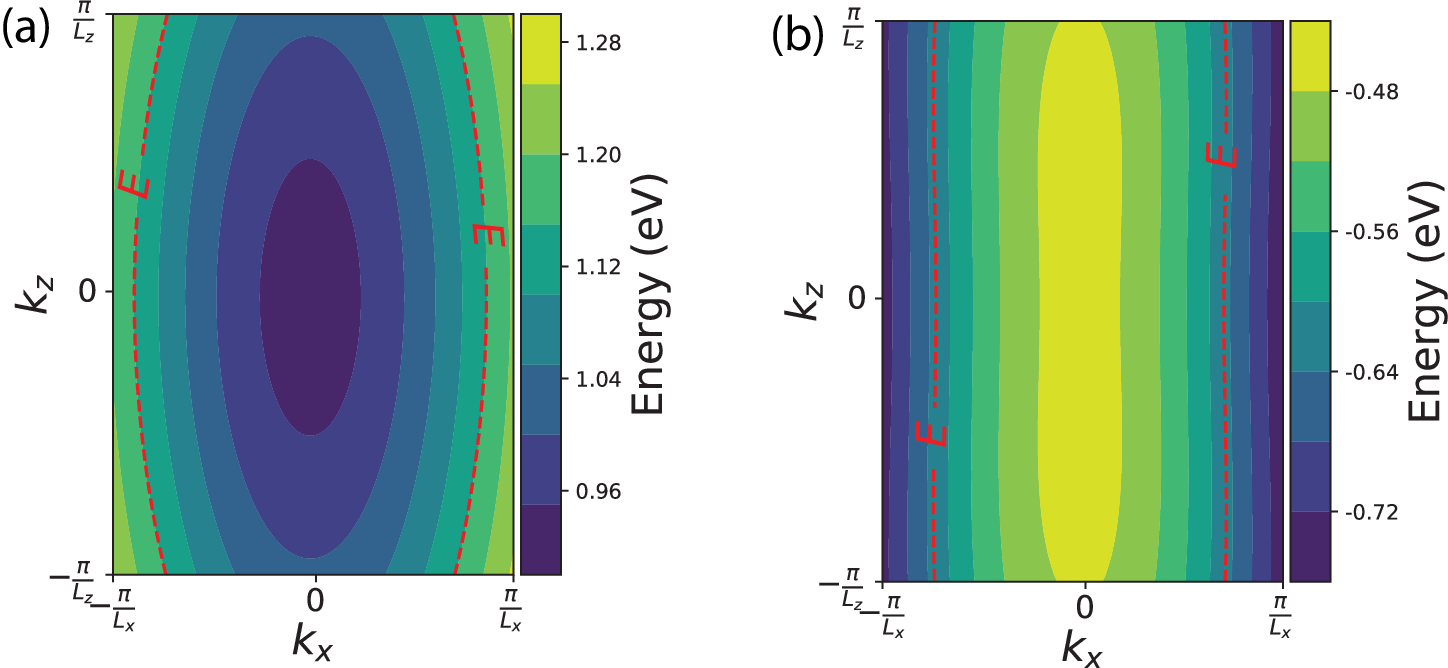}
\caption{(a) Contour plot of the bottom conduction band of a black
  phosphorus monolayer corresponding to an orthorhombic (rectangular)
  unit cell of dimensions $L_x$=2.77 nm and $L_z$=1.98 nm. In the
  transistor configuration, $k_x$ is aligned with the transport
  direction, $k_z$ with the direction assumed periodic. The dashed red
  line indicates the Fermi level iso-energy. The effective mass along
  $k_x$, $m_{\Gamma-X}$, is equal to 0.16 $m_0$, along $k_z$,
  $m_{\Gamma-Y}$=1.2 $m_0$ (b) Same as (a), but for the top valence
  band of black phosphorus ($m_{\Gamma-X}$=0.14 $m_0$,
  $m_{\Gamma-Y}$=4.46 $m_0$). Adapted from \cite{klinkert}.}
\label{fig:anis}
\end{figure}

Among all 2-D materials that were simulated, black phosphorus stands
out as it displays the largest ``$n$-type $I_{ON}$ vs. $p$-type
$I_{ON}$'' combination, when the transport direction of the 
FET is aligned with the $\Gamma$-$X$ crystal axis of BP. The
conduction and valence band anisotropy of this monolayer (see
Fig.~\ref{fig:anis}) is at the origin of the high current
densities. Its transport effective mass, $m_{trans}$, is equal to 0.16
$m_0$ for electrons (0.14 for holes), whereas its density-of-states
counterpart, $m_{DOS}$, amounts to 0.42 $m_0$ (0.82 for holes). Other
2-D materials benefit from anisotropic bandstructures, which is the
reason why they deliver $I_{ON}$'s larger than 3 mA/$\mu$m, as can be
generally seen in Fig.~\ref{fig:ion}(b-c). This is the case of the
electrons and holes in Ag$_2$N$_6$, As$_8$Ge$_4$, or As$_8$Si$_4$, and
of the holes in I$_4$O$_4$Sc$_4$, for example. Their band extrema have
an ellipsoid shape. In fact, almost all 13 best components are
characterized by a $m_{trans}$ lower than their $m_{DOS}$, i.e. they
are situated above the dashed black lines that correspond to materials
with an isotropic bandstructure in Fig.~\ref{fig:ion}(b-c). It should
be noted that $m_{trans}$ and $m_{DOS}$ were not directly extracted
from the bandstructure of the 2-D materials, but by calculating
their charge and current densities with analytical equations,
as in \cite{klinkert}.

The question that arises with materials having an anisotropic
conduction band minimum and/or valence maximum is ``what happens if
the direction along which the electrical current flows is not
perfectly aligned with the most suitable crystal axis?''. For example,
it is well-known in BP that transport along the $\Gamma$-$Y$ axis is
(much) less efficient than along $\Gamma$-$X$ (\cite{anisotropy}). Using
the proposed MLWF+NEGF approach, we found in \cite{klinkert2} that
orientation misalignments up to $50^{\circ}$ from the ideal case do
not significantly alter the ON-state current of BP transistors, with
almost negligible performance loss up to a misalignment angle of
$20^{\circ}$. This means that $I_{ON}$ does not linearly decrease as a
function of the misalignment angle, but rather first stays constant up
to $20^{\circ}$, slightly decreases up to $50^{\circ}$, and finally
rapidly drops. This behavior, which occurs both in the ballistic limit
of transport and in the presence of electron/hole-phonon and charged
impurity scattering, can be explained by considering the
angle-dependent value of the $m_{trans}$ and $m_{DOS}$ effective
masses.

It is important to realize that the impact of misalignment angles that
was discussed above for black phosphorus transistors can be
generalized to any 2-D materials with an anisotropic bandstructure. If
the $m_{\Gamma-X}/m_{\Gamma-Y}$ ratio of the effective masses extracted
from the bandstructure along the two main axes of the Brillouin Zone
is smaller than 0.1, i.e. if $m_{\Gamma-Y}\geq$10$m_{\Gamma-X}$ in
Fig.~\ref{fig:anis}, the ON-state current only marginally decreases up
to $\delta\leq20^{\circ}$, by $25\%$ if the misalignment is pushed to
$40^{\circ}$. It is clear that the magnitude of the ON-state current
still depends on the DOS of each 2D material. Nevertheless, a region where
the current is almost insensitive to the misalignment angle can be
expected in all cases, as suggested \cite{klinkert2}. 

\begin{figure}[ptbh]
\centering
\includegraphics[width=\linewidth]{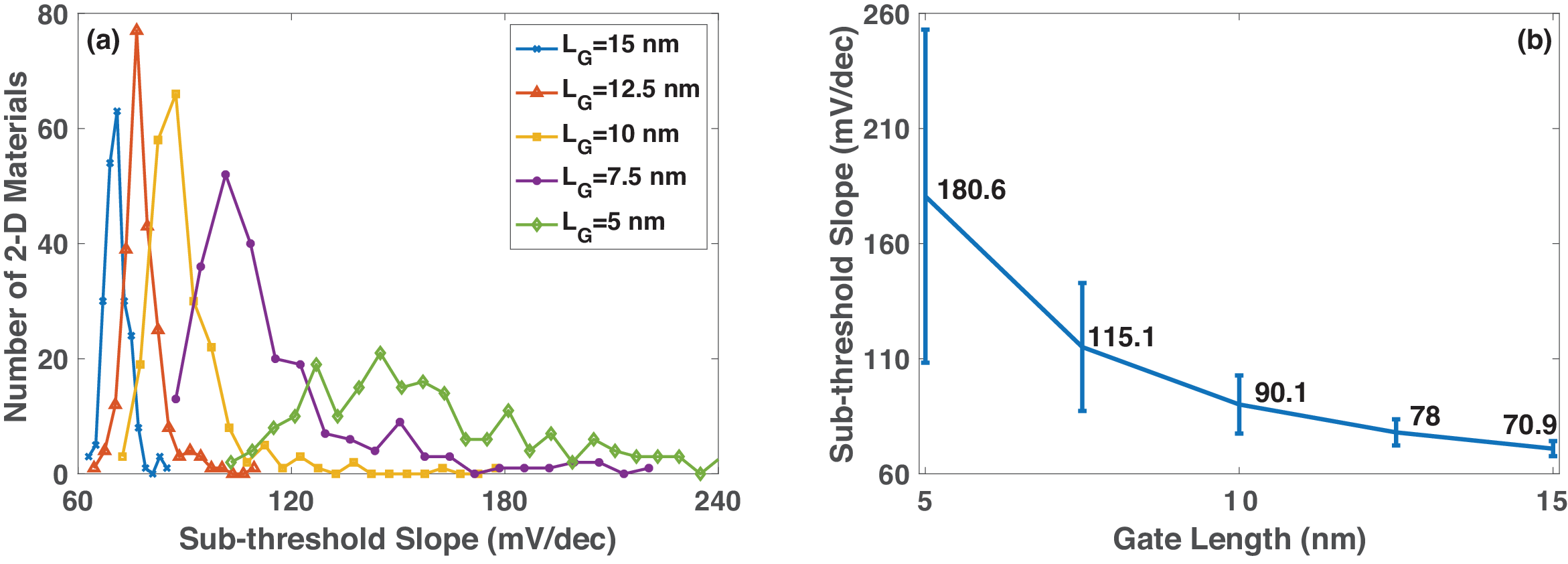}
\caption{(a) Histogram of the sub-threshold slope of all investigated
  $n$- and $p$-type FETs at a gate length of $L_G$=15, 12.5, 10, 7.5,
  and 5 nm. (b) Average $SS$ as a function of $L_G$. The error bars
  indicate the standard deviation. Adapted from \cite{klinkert}.}
\label{fig:ss}
\end{figure}

If 2-D materials should replace Si as the channel of future
field-effect transistors, they should be used during several
consecutive technology nodes. Consequently, it should be possible
to scale their gate length and still get a performance superior to
that of Si FinFETs or their potential successors, gate-all-around
nano-sheets, see \cite{nanosheet}. Thus, in Fig.~\ref{fig:ss}, we show
the sub-threshold slope of all simulated 2-D FETs as histograms for the
gate lengths $L_G$=15, 12.5, 10, 7.5, and 5 nm. The corresponding
average values and standard deviations are provided in the
right-hand-side of the plot. While $SS$ slowly increases when $L_G$
shrinks from 15 down to 10 nm, it explodes below 10 nm, as does the
variations among the different 2-D materials. From these results, it
does not seem feasible to scale single-gate 2-D FETs below 10 nm. By
adding a second, symmetric gate at the bottom of the structure in
Fig.~\ref{fig:mosfet} an improvement of the transistor scalability can
be expected. The double-gate architecture is probably the only viable
path for 2-D materials to compete with Si FETs. Still,
other technology issues remain to be solved, as detailed in the next
section. At the same time, new logic switch opportunities could emerge
for 2-D materials.

\begin{table}
  \centering
\begin{tabular}{|l|cccc|cccc|}
\cline{2-9}
\multicolumn{1}{c|}{} & \multicolumn{4}{c|}{$n$-FET} & \multicolumn{4}{c|}{$p$-FET}\\
\cline{2-9}         
\multicolumn{1}{c|}{} & \makecell{$I_{ON}$\\(mA/$\mu$m)} & \makecell{$m_{tran}$\\($m_0$)} 
& \makecell{$m_{DOS}$\\($m_0$)}  & \makecell{$SS$\\(mV/dec)} &
\makecell{$I_{ON}$\\(mA/$\mu$m)} & \makecell{$m_{trans}$\\($m_0$)} 
& \makecell{$m_{DOS}$\\($m_0$)} & \makecell{$SS$\\(mV/dec)}  \\
\hline
  Ag$_2$N$_6$       &   3.91  & 0.31  & 0.56 &  67  &  3  & 0.42  & 0.65 &  66.6  \\
\hline
  As$_2$            &   4.17  & 0.24  & 1.67   &  67.4  &  3.77  & 0.34   & 0.56   &  69.7  \\
\hline
  As$_4$            &   4.44  & 0.25  & 0.61   &  68.4  &  5.18  & 0.1   & 0.35   &  73.8 \\
\hline
  Ge$_2$S$_2$       &   4.45  & 0.24  & 0.63   &  67.8  &  4.08  &  0.23  & 0.7   &  65.8  \\
\hline
  Ge$_2$Se$_2$      &   3.4  & 0.33  & 0.5   &  69.1  &  4.38  &  0.35  & 0.44  &  68  \\
\hline
  HfS$_2$           &   3.27  & 0.51  & 2.33   &  69.1  &  3.59  & 0.35   & 0.66   &  68.8  \\
\hline
  O$_6$Sb$_4$       &   4.47  & 0.18 & 2.04   &  67.6  &  3.07  & 0.35   & 0.73   &  62.6  \\
\hline
  BP             &   5.01 & 0.16   & 0.42  &  69.5  &  5.36  & 0.14   & 0.82   &  70.4 \\
\hline
  Sb$_2$            &   5.11  & 0.23  & 0.55   &  67.5  &  3.8  & 0.3   & 0.52   &  70.1  \\
\hline
  Si$_2$H$_2$       &   4.46  & 0.29  &  1.74  &  69.5  &  3.2  &  0.43  & 0.71   &  69.5  \\
\hline
  Ti$_2$Br$_2$N$_2$ &   3.49  & 0.18  & 0.23   &  71.6  &  3.25  & 0.18  & 0.24   &  75.8  \\
\hline
  Ti$_2$N$_2$Cl$_2$ &   3.44  & 0.22  & 0.29  &  69.9  &  3.48  & 0.19   & 0.24   &  70.4  \\
\hline
  ZrS$_2$           &   3.24  & 0.59  & 2.45   &  66.4  &  3.37  & 0.37   & 0.7   &  68.5  \\
\hline
\end{tabular}
\caption{ON-state current ($I_{ON}$), transport effective mass
  ($m_{trans}$), density-of-states effective mass ($m_{DOS}$), and
  sub-threshold slope ($SS$) of the 13 2-D single-gate FETs with
  $I_{ON}>$3 mA/$\mu$m in both their $n$- and $p$-type configurations  
  at $L_G$=15 nm. If more than one transport direction satisfies the
  ON-state current condition for a given 2-D material, only the best
  one is reported here. Adapted from \cite{klinkert}.}
\label{tab:2}
\end{table}

\section{Challenges and Opportunities}\label{sec:chall}

\begin{figure}[ptbh]
\centering
\includegraphics[width=\linewidth]{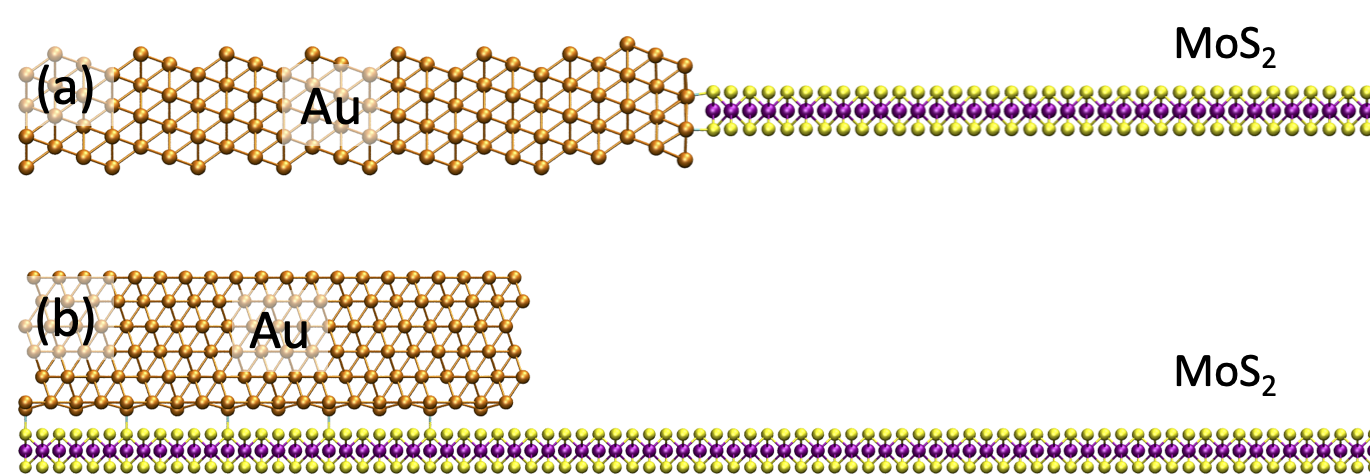}
\caption{(a) Schematic view of an Au-MoS$_2$ structure with a side
  contact configuration, as simulated with the MLWF+NEGF approach
  introduced in Section \ref{sec:mod}. (b) Same as (a), but for an Au
  top contact.}
\label{fig:contact}
\end{figure}

\subsection{Electrical Contacts between Metals and 2-D Monolayers}

One of the key challenges 2-D materials are facing is their contacting
with metallic electrodes. Different approaches can be used to inject
electrons into monolayers, the most common ones being top (\cite{kis}),
side (\cite{side}), and phase-engineered contacts (\cite{kappera}).
The former two are plotted in Fig.~\ref{fig:contact}. Besides these geometries,
the choice of the metal (\cite{das}), the introduction of an interfacial
layer between the metal and semiconductor (\cite{kim2}), or the doping of
the channel (\cite{wallace2}) represent additional design
options. Typically, all these contact configurations are characterized
by resistances in the k$\Omega\cdot\mu$m range (\cite{franklin2}) instead
of 150 to 200 $\Omega\cdot\mu$m as in Si FinFETs, see \cite{irds}. It
should nevertheless be mentioned that phase-engineered or
nickel-etched graphene electrodes, as in \cite{qw0}, can give lower
resistance values, in the order of 200 $\Omega\cdot\mu$m, but for
multilayer, not monolayer MoS$_2$.

A question that remains open concerns the transfer of electrons from
the metallic contact to the semiconductor channel, especially in the
case of top-contact architectures. The transfer length $L_T$ measures
the average distance that an electron needs to completely leave the
metal electrode and enter the semiconducting channel. Two different
scenarios are possible, one labeled ``edge process'' (electrons flow
through the metal up to the edge of the metal-semiconductor interface,
$L_T$ is close to 0), the other one ``area-dependent process''
(electrons are gradually transferred from the metal to the semiconductor,
$L_T\gg$0). For example, in \cite{large_transfer_length}, a transfer
length $L_T\simeq 600$ nm was found for a monolayer MoS$_2$ with titanium
contacts, while in \cite{English} Au electrodes on top of bilayer
MoS$_2$ led to low contact resistances $R_C$=740 $\Omega\cdot\mu$m and
a short $L_T$ of roughly 30 nm, i.e. a nearly edge process.

On the theoretical side, different explanations for such a behavior
have been proposed. Relying on DFT, \cite{Kang2} came to the conclusion
that the MoS$_2$ layer below the metallic contact metallizes, which
creates an area-dependent injection of electrons. Through device
simulations performed in the effective mass approximation,
\cite{meff_contact} determined that the transfer length depends on the
number of layers composing the 2-D material, going from an edge process in
monolayers to an area-dependent one in multi-layer compounds. Using
our \textit{ab initio} quantum transport approach, we managed to
reconcile these theories. If the metal-semiconductor interface is
perfectly clean, the transmission of electrons from the electrode to
the channel tends to be edge-dependent in monolayers, while the
transfer length increases if an interfacial layer is present between
both materials (\cite{szabo4}). While the exact nature of the transfer
process is not yet fully understood, it is clear that $L_T$ should be
as small as possible to make the 2-D technology fully scalable. Recently,
\cite{smets} demonstrated few-layer MoS$_2$ FETs with
13-nm-long top contacts and still very good performance in terms of
ON-state current, sub-threshold slope, and contact resistance. These 
devices have therefore an ultra-short $L_T$, which goes exactly in the
right direction. Note that side contacts are intrinsically more
scalable, but they do not currently offer the same performance as top
ones, as illustrated in \cite{jain}.

\begin{figure}[ptbh]
\centering
\includegraphics[width=\linewidth]{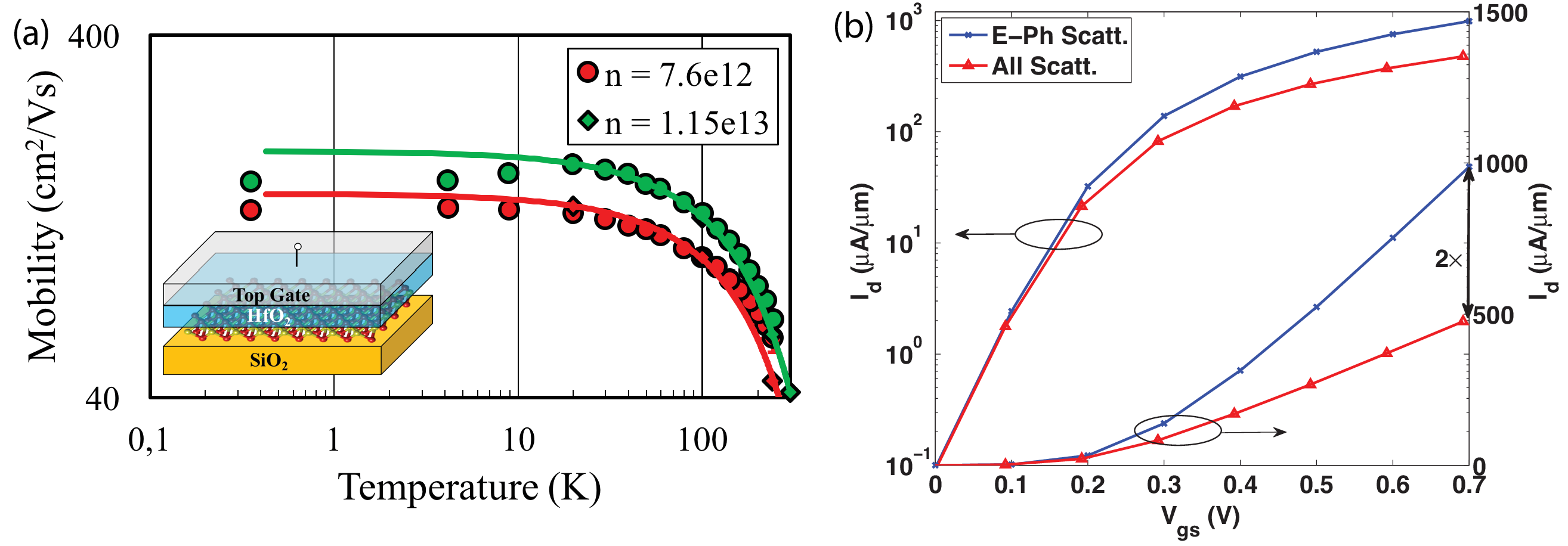}
\caption{(a) Temperature-dependent mobility of monolayer MoS$_2$
  embedded between two oxide layers (see inset) at two different
  carrier densities. Symbols refer to the experimental data of
  \cite{kisnatmat}, lines to simulation results. Electron-phonon,
  charged impurity (CI), and surface optical phonon (SOP) scattering
  are included. (b) Transfer characteristics $I_D$-$V_{GS}$
  at $V_{DS}$=0.7 V of a MoS$_2$ 2-D FET with the same single-gate
  structure as in Fig.~\ref{fig:mos2} and $L_G$=15 nm. The current
  with electron-phonon interactions only (blue line with crosses) and
  with additionally CI and SOP scattering (red lines with triangles)
  are reported. Adapted from \cite{lee}.}
\label{fig:mob}
\end{figure}

\subsection{2-D Mobility Limiting Factors}

\textit{Ab initio} calculations have been widely used to predict the
phonon-limited mobility of 2-D materials, with a strong focus on
TMDs. One of the first such calculations was done in 2012 by
\cite{kaasbjerg} for monolayer MoS$_2$. The electron-phonon scattering
rates obtained from DFT served as inputs to fit deformation
potentials that were then used in a standard linearized Boltzmann
Transport Equation (LBTE) solver. A mobility of 410 cm$^2$/Vs at room
temperature was returned by this approach. It is much larger than what
has been so far measured experimentally, e.g. 63 cm$^2$/Vs at 240 K in
\cite{kisnatmat} or 35.7$\pm$2.6 cm$^2$/Vs at 300 K in \cite{smithe},
depending on the dielectric environment. More recent MoS$_2$ mobility
calculations based on fully \textit{ab initio} electron-phonon
scattering rates revealed values much closer to experiments, 150
cm$^2$/Vs in \cite{wuli} or 144 cm$^2$/Vs in \cite{sohier}. However,
other effects such as charged impurity scattering (CIS) (\cite{zong}) or
surface optical phonons (SOP) (\cite{nma}) could also play a critical
role and bring the mobility closer to measurements. 

In Fig.~\ref{fig:mob}(a), we show mobility results for single-layer
MoS$_2$ that were computed with our in-house LBTE solver described in
\cite{rrhyner}, including \textit{ab initio} electron-phonon
interactions as well as CIS and SOP. The only parameter that was
neither calculated nor taken from the literature is the concentration
of charged impurities, $n_{imp}$. This quantity was used as fitting
parameter to best reproduce the experimental data of \cite{kisnatmat},
which was achieved with $n_{imp}$=2.5$\times$10$^{12}$ cm$^{-2}$. An excellent
agreement between simulations and experiments is obtained for two
different electron concentrations, 7.6$\times$10$^{12}$ and
1.15$\times$10$^{13}$ cm$^{-2}$.

Getting the mobility gives a lot of information about the transport
properties of a material, but being able to determine the influence of
this figure of merit on the ``current vs. voltage'' characteristics of
a device is equally important. To do that, we used our LBTE solver
to calibrate the magnitude of the scattering mechanisms in our MLWF+NEGF
QT simulator, OMEN (\cite{lee}). This step is necessary as several
approximations to the scattering self-energies must be applied in the
NEGF formalism. To compensate for them, each scattering rate
(electron-phonon, CIS, and SOP) can be scaled by a different factor so
that both LBTE and QT produce the same mobility. This is what we did
for a MoS$_2$ monolayer before simulating the transfer characteristics
of a device similar to the one in Fig.~\ref{fig:mos2}(a). Results are
presented in Fig.~\ref{fig:mob}(b), where the currents with
electron-phonon scattering only and with additionally CIS and SOP are
compared to each other. It can be seen that CIS and SOP, together, are
responsible for a reduction of the ON-state current by a factor 2, as
compared to the case without them so that $I_{ON}$ does not exceed 500
$\mu$A/$\mu$m at a gate length $L_G$=15 nm. This finding indicates
that there is room for improvement. The impurity concentration might
indeed be decreased by improving the 2-D crystal quality and its
interface with the dielectric environment, while SOP might be
minimized through substrate engineering. Furthermore, strain might
have a beneficial impact on the electronic properties of MoS$_2$, as
theoretically demonstrated in \cite{sohier_2019} for other 2-D
materials. Overall, there are certainly multiple paths to push the
ON-state current of MoS$_2$ above 1 mA/$\mu$m.

\begin{figure}[ptbh]
\centering
\includegraphics[width=\linewidth]{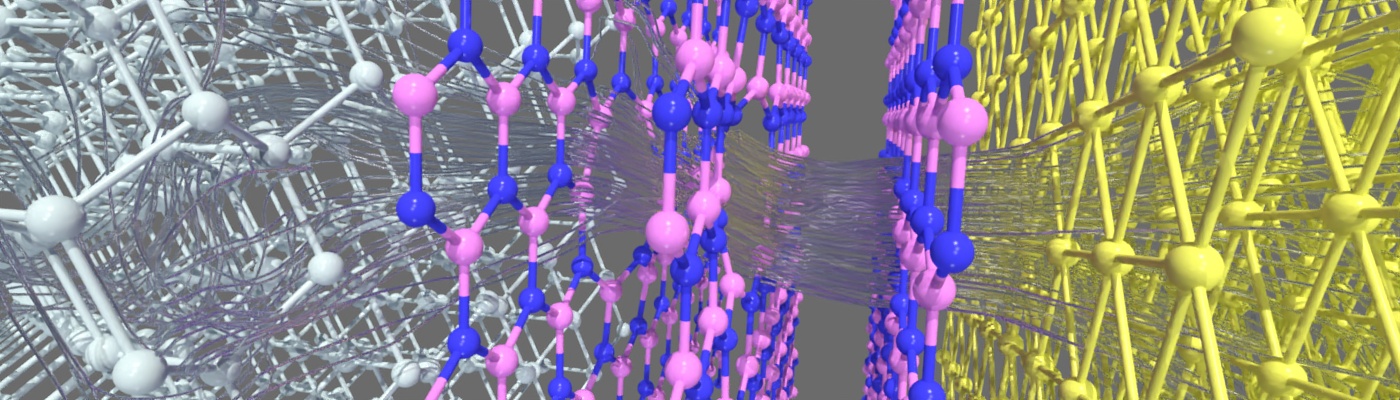}
\caption{Leakage current trajectories through a metal (Au) - oxide
  (hexagonal boron nitride) - semiconductor (Si) structure. The oxide
  layer is made of 3 h-BN layers, which corresponds to a physical
  thickness of $\sim$1 nm and an equivalent oxide thickness EOT=0.76  
  nm ($\epsilon_R$=5). The h-BN insulator contains a defect in its
  central layer. Credit: Dr. Jean Favre (CSCS).}
\label{fig:hbn}
\end{figure}

\subsection{2-D Oxides}

What has made silicon such an attractive channel material for
transistors is the existence of a native oxide, SiO$_2$, that can be
used to separate it from gate contacts or from another
dielectric layer. Although not perfect, the Si-SiO$_2$ interface
contains a low defect density, contrary to what is found if SiO$_2$,
Al$_2$O$_3$, or HfO$_2$ is deposited on a TMD monolayer, see \cite{grasser_2020}.
There are few exceptions such as ZrO$_2$ on ZrSe$_2$ or HfO$_2$ on
HfSe$_2$: although large ON/OFF current ratios were demonstrated, the extracted
carrier mobilities remained small and the \textit{I-V} characteristics
still exhibited an hysteresis (\cite{mleczko}) when sweeping the gate
voltage back and forth. Such a behavior is usually induced by the
presence of interface traps that are charged and discharged as the
gate potential varies.

Another approach consists of placing a 2-D oxide on a 2-D channel
material. The most common 2-D insulator is hexagonal boron nitride
(hBN). For example, encapsulating MoS$_2$ between hBN layers has been
shown to produce higher mobility values, both in mono- and few-layer
configurations (\cite{hbn_mobility}). Using hBN in transistor
applications might however not be optimal as the relative dielectric
permittivity, $\epsilon_R$, of this materials is in the order of 5. To
reach an EOT of 1 nm or less, a hBN thickness of $<$1.3 nm is needed,
which corresponds to 3 to 4 layers. We used our \textit{ab
  initio} QT simulator to determine what the implications of such thin
oxides might be for gate leakage currents (\cite{knobloch}). As testbed,
a Au-hBN-Si metal-oxide-semiconductor (MOS) capacitor was constructed
at the atomic level and the current that flows through it computed as
a function of the applied voltage. It was found that 3 and 4 layers of
hBN are not sufficient to satisfy the IRDS requirements. The situation
dramatically worsens if a defect is present in the hBN
dielectric. Bridges can form between adjacent hBN layers, which
locally increases the leakage current, as can be seen in
Fig.~\ref{fig:hbn}. All current trajectories converge towards the
defect location.

Other 2-D oxides have been deposited on TMD channels, e.g. CaF$_2$
in \cite{grasser_2019}. The advantages of such crystals are that they
have a perfectly ordered structure, contrary to SiO$_2$ or HfO$_2$, which
are amorphous, and that they form a quasi van der Waals interface
with 2-D channels. Furthermore, the dielectric constant of CaF$_2$
is higher than that of hBN such that 2 nm of this oxide yields an EOT 
of roughly 1 nm. Given the fact that the 2-D material space includes
more than 1,800 compounds (\cite{mounet}), it can be expected that
several other 2-D oxides might be exfoliated in the future and that
some of them are competitive with HfO$_2$ in terms of $\epsilon_R$ and
band offsets, while preserving clean oxide-semiconductor interface.  

\begin{figure}[ptbh]
\centering
\includegraphics[width=\linewidth]{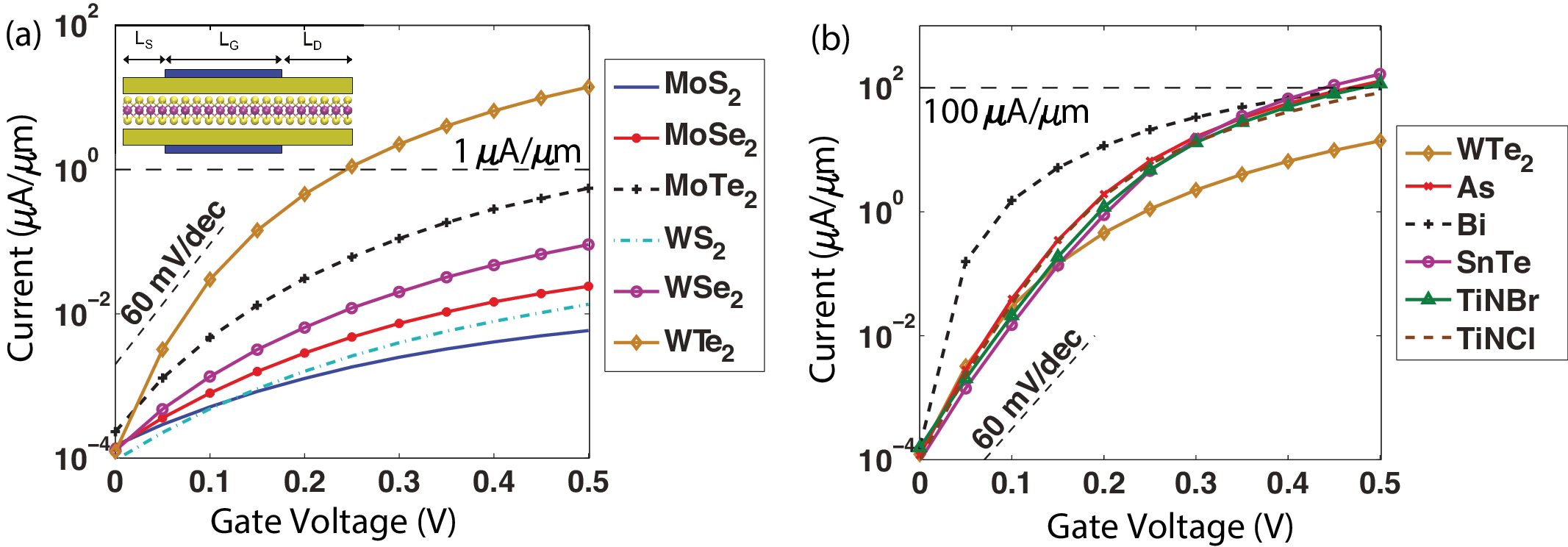}
\caption{(a) Transfer characteristics $I_D$-$V_{GS}$ at $V_{DS}$=0.5 V
  of band-to-band tunneling field-effect transistors (TFETs) made of
  conventional TMDs (+WTe$_2$) with a double-gate configuration (see
  inset). The gate length $L_G$ is set to 40 nm, the source and drain
  extensions to $L_S$=13 nm (acceptor doping concentration $N_A$=5$\times$10$^{13}$
  cm$^{-2}$) and $L_D$=26 nm (donor doping concentration $N_D$=5$\times$10$^{12}$
  cm$^{-2}$), respectively. The 60 mV/dec slope and the 1
  $\mu$A/$\mu$m current level are indicated by dashed black lines. (b)
  Same as (a), but for novel 2-D materials from the database of
  \cite{mounet}. Adapted from \cite{szabo3}.}
\label{fig:btbt}
\end{figure}

\subsection{Advanced Logic Concepts}

2-D materials do not only face challenges, they also offer
opportunities in advanced logic applications. Their excellent
electrostatic control properties are particularly appealing for the
realization of band-to-band tunneling field-effect transistors
(TFETs). An entire chapter of this book is dedicated to these logic
concepts that could theoretically exhibit a sub-threshold slope below
60 mV/dec at room temperature (\cite{riel_11}). Such a feature enables
for a drastic reduction of the supply voltage and therefore of the
power consumption of integrated circuits. However, due to the
tunneling nature of the injection mechanism, TFETs tend to suffer from
very low ON-state currents. Typically, a large band gap is
needed to ensure a steep $SS$ and a low $I_{OFF}$, while a small band
gap is necessary to boost $I_{ON}$. This dichotomy can be partly
addressed through the usage of heterojunctions, as demonstrated in
\cite{lund}. 

Through device simulation, it has been shown in
\cite{2dtfets} and \cite{metal_dichalcogenide_tfet1} that conventional
TMD monolayers are probably not the best TFET candidates as their 
large band gaps do not allow the ON-state current to reach large
values. This behavior is confirmed in Fig.~\ref{fig:btbt}(a): if the
OFF-state current is fixed to $\sim$0.1 nA/$\mu$m and the supply
voltage $V_{DD}$ to 0.5 V, $I_{ON}$ does not exceed 1 $\mu$A/$\mu$m,
except for WTe$_2$, a TMD that is rather difficult to stabilize in the 2H
phase. Even worse, the $SS$ is superior to 60 mV/dec for several of
the considered TMDs. This can be explained by the fact that even in
the OFF-state, the tunneling channel is already open, i.e. the
conduction band below the gate contact has already been pushed below
the valence band of the source, see \cite{szabo3}. Once this condition
is satisfied, the $I_{ON}$ increase with respect $V_{GS}$ slows down and
becomes almost linear instead of exponential.

There are different solutions to obtain a better TFET performance, for
example by combining different TMDs and forming van der Waals
heterostructures. The benefit of such approaches has been highlighted
both theoretically in \cite{metal_dichalcogenide_tfet2},
\cite{szabo2}, or \cite{pala} and experimentally in
\cite{ge_mos2_tfet}, \cite{balaji}, or \cite{oliva} combining
different TMDs together or one TMD with another material,
e.g. germanium. Alternatively, the huge variety of properties
encountered in 2-D materials (see \cite{mounet}) can be taken
advantage of to identify compounds with low band gaps, compatible with
a high ON-state current. With the help of our MLWF+NEGF solver, we
have simulated the electrical behavior of relevant examples. Their
transfer characteristics are displayed in Fig.~\ref{fig:btbt}(b). All
have $I_{ON}$'s in the order of 100 $\mu$A/$\mu$m at $V_{DD}$=0.5 V
and $I_{OFF}$=0.1 nA/$\mu$m, which is 100$\times$ larger than most
TMDs. At the same time, a steep $SS$ is obtained. Further
investigations would be needed to screen the available design space,
which may contain many more 2-D materials with band gaps comprised
between 0.5 and 1 eV.

\section{Conclusion and Outlook}\label{sec:conc}

In this Chapter, the potential of 2-D materials as field-effect
transistors has been discussed from a modeling perspective, starting
from the key features of monolayers. The importance of being able to
simulate their electrical characteristics has been introduced. Among
all possible approaches, the combination of plane-wave density
functional theory, maximally localized Wannier functions, and quantum
transport has been selected for its versatility and accuracy. By
applying it, it has been revealed that transition metal
dichalcogenides cannot currently provide a switching performance that
is comparable to that of Si FinFETs. They suffer from technical
difficulties that will probably disappear with time, such as 
their electrical contacting, as well as from inherent deficiencies,
e.g. relative large effective masses that negatively impact their
carrier mobilities. We have shown theoretically that novel 2-D
materials represent an attractive alternative to TMDs with excellent
figures of merits as logic switches. It remains to confirm
experimentally the predicted performance, which requires first to
isolate the desired mono- or few-layer components.  

\section{Acknowledgment}
The work presented in this Chapter was supported by ETH Zurich (grant
ETH-32 15-1) and by the Swiss National Science Foundation (SNSF) under
grant no. 200021\_175479 (ABIME) and under the NCCR MARVEL. We
acknowledge PRACE for awarding us access to Piz Daint at CSCS under
Project pr28, PRACE for the allocated computational resources on
Marconi at CINECA under Project 2016163963, and CSCS for Project
s876.

\bibliographystyle{IEEEtran}
\bibliography{paper}

\vfill\pagebreak
\
\thispagestyle{empty}
\end{document}